\documentclass[prb,twocolumn,showpacs,amssymb]{revtex4}

\usepackage{graphicx}
\usepackage{bm}
\usepackage{amssymb,amsmath, mathrsfs}
\usepackage{color}
\def\be{\begin{equation}}
\def\ee{\end{equation}}
\def\bea{\begin{eqnarray}}
\def\eea{\end{eqnarray}}

\newcommand{\gs}{{\gtrless}}

\begin{document}

\title{Theory of non-equilibrium electronic Mach-Zehnder interferometer}
\author{
Martin Schneider,$^{1,2}$ Dmitry A. Bagrets,$^{3,4}$ and Alexander
D.~Mirlin$^{1,3,5,6}$ 
}
\affiliation{
$^{1}$Institut f\"ur Theorie der \!Kondensierten \!Materie, Karlsruhe
Institute of Technology, 76128 Karlsruhe, Germany\\
$^{2}$ Dahlem Center for Complex Quantum Systems and Institut f\"ur Theoretische Physik,
Freie Universit\"at Berlin, 14195 Berlin, Germany \\ 
$^{3}$Institut f\"ur Nanotechnologie, Karlsruhe Institute of
Technology, 76021 Karlsruhe, Germany\\
$^{4}$Institut f\"ur Theoretische Physik, Universit\"at zu K\"oln, 
Z\"ulpicher Str. 77, 50937 K\"oln, Germany\\
$^{5}$DFG Center for Functional Nanostructures, Karlsruhe Institute of
Technology, 76128 Karlsruhe, Germany\\  
$^{6}$Petersburg Nuclear Physics Institute, 188300 St.~Petersburg, Russia
}

\date{\today}

\begin{abstract}
We develop a theoretical description of interaction-induced phenomena
in an electronic Mach-Zehnder interferometer formed by integer quantum
Hall edge states (with $\nu =1$ and 2 channels) out of equilibrium. Using
the non-equilibrium functional bosonization framework, we derive an
effective action which contains all the physics of the problem. We
apply the theory to the model of a short-range interaction and to a
more realistic case of long-range Coulomb interaction. The theory
takes into account interaction-induced effects of dispersion of
plasmons, charging, and decoherence. In the case of long-range
interaction we find a good agreement between our theoretical results
for the visibility of Aharonov-Bohm oscillations and experimental data. 
\end{abstract}

\maketitle

\section{Introduction}
\label{s1}

Many recent experiments studied transport through an electronic analog
of Mach-Zehnder interferometer (MZI) built on edge states in the quantum
Hall regime \cite{heiblum03,heiblum06,heiblum07,heiblum07a,roche07,roche08,roche09,
strunk07,strunk08,strunk10,schoenenberger}. These
experiments show strong Aharonov-Bohm oscillations, which is a
manifestation of quantum interference of electrons propagating along
the arms of the interferometer. One of remarkable experimental
observation is a lobe-type structure in the dependence of visibility on bias voltage. 
More precisely, the visibility strongly depends on voltage, showing decaying
oscillations (``lobes'') characterized by certain energy scale.
Such structure can not be explained within a model of non-interacting electrons 
(which would predict a constant visibility) and thus results from
the electron-electron  interaction. Therefore, the experiments on Mach-Zehnder
interferometers exhibit the physics  
resulting from an interplay of a quantum interference and the Coulomb
interaction under strongly 
non-equilibrium conditions. Development of a theory of such phenomena
is a challenging task. 

Various aspects of the relevant physics have been addressed in earlier
works. Dephasing of quantum interference in Aharonov-Bohm rings,
interferometers and related phenomena have been studied in many works
for equilibrium 
\cite{aronov87,buettiker,LeHur,marquardt04,ludwig04,gornyi05,dmitriev10} 
and non-equilibrium 
\cite{neder07,gutman08,gutman10,NgoDinh10}
setups. 
The importance of electron-electron interaction for the emergence of
the lobe structure has been emphasized in
Refs.~\onlinecite{sukhorukov07,Levkivskyi08}. In Ref.~\onlinecite{Chalker07} 
the influence of long-range ($1/r$) character of the Coulomb
interaction (leading to a dispersion of the plasmon mode) was analyzed. 
In the works \onlinecite{neder08,youn08,kovrizhin09} a simplified model with
a quantum-dot-type treatment of the interaction was solved. 

The goal of the present article is to present a systematic theory of
transport in a non-equilibrium quantum Hall Mach-Zehnder
interferometer and to confront its predictions with the experiment. 
This theory  
formulated in terms of functional-bosonization Keldysh action in
Sec.~\ref{s2} takes into account all effects of the electron-electron
interaction, including formation and characteristics of the lobe
structure, as well as the dephasing. In Sec.~\ref{s3} we apply the theory to the
model of short-range interaction. In this way we reproduce the earlier
results on the lobe structure  \cite{Levkivskyi08} and also find the
suppression of the interference signal due to dephasing. While showing
some similarity to experimental observations, the results of the
short-range-interaction model 
contradict to the experiment in several crucial aspects. 
This motivates us to explore the more realistic case of $1/r$ Coulomb
interaction in Sec.~\ref{s4}.  Using our general formalism, we analyze the
cases of $\nu=1$ and $\nu=2$ edge modes. The obtained parametric
dependences of the interference signal are in good agreement with
experiments, although in the case of $\nu=1$ mode some discrepancies
remain. We summarize our findings and discuss further research
directions in Sec.~\ref{s5}.

\section{General framework}
\label{s2}
\subsection{Model and the functional bosonization}
\label{s2.1}

\begin{figure}[b]
\includegraphics[width=3.5in]{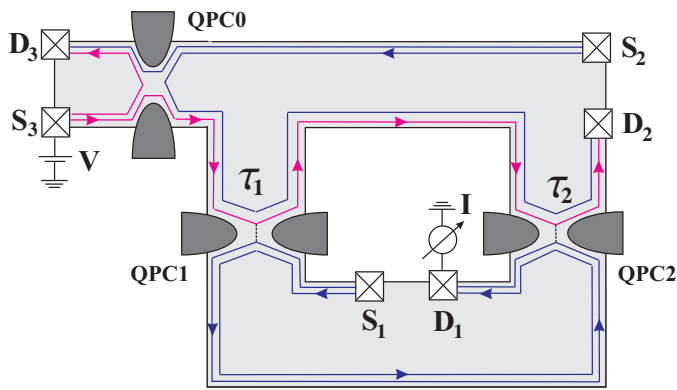}
\caption{Scheme of an electronic Mach-Zehnder interferometer
built on quantum-Hall edge states at filling factor $\nu=2$.
Quantum point contact (QPC1 and QPC2) characterized by transmission amplitudes
$\tau_{1(2)}$ are used to partially mix the outer edge channels. All
Ohmic contacts are grounded, except for the source terminal $S_3$
which is kept at voltage $V$. The current is 
measured in the drain terminal $D_1$. The QPC0 is pinched in such a
way that the inner channel is completely reflected while the outer
one is fully transmitted.}
\label{MZI_scheme}
\end{figure}

\begin{figure}[t]
\includegraphics[width=2.3in]{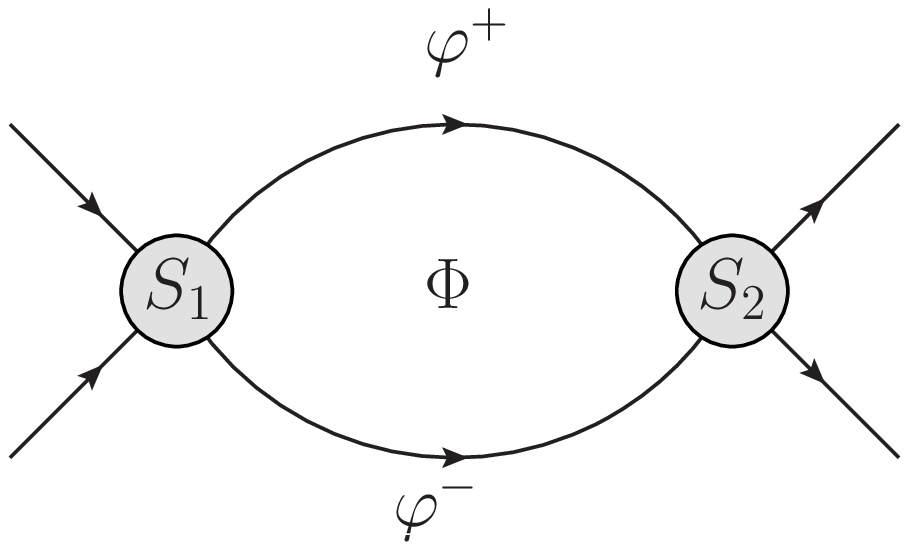}
\includegraphics[width=2.3in]{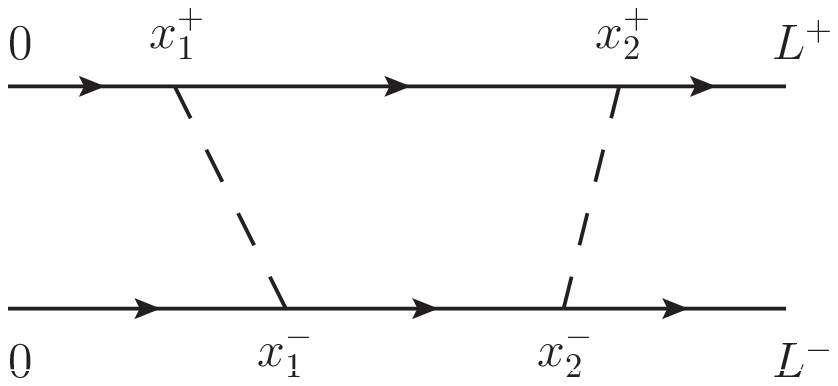}
\caption{{\it Top:} Schematics of a Mach-Zehnder interferometer. Two
  quantum point contacts are characterized by scattering matrices
  $S_1$ and $S_2$. The electron-electron interaction is decoupled via
  a Hubbard-Stratonovich fields $\varphi^+$ and $\varphi^-$ on two arms of
  the interferometer. {\it Bottom:} Coordinates on two arms of the
  interferometer.}
\vspace{-2mm}
\label{fig:MZI}
\end{figure}

We consider a theoretical model of the Mach-Zehnder interferometer,
realized with edge states in the quantum Hall regime at filling fraction  
$\nu=1$ and $\nu=2$, as schematically shown in
Figs.~\ref{MZI_scheme} and \ref{fig:MZI}. The outer channels propagating along 
different arms (index: $\pm$) of the interferometer are coupled by means of two
quantum point contacts (QPCs), located at points $x_1^{\pm}$  
and $x_2^{\pm}$.  Each QPC can be generally described by the unitary
$2\times 2$ scattering matrix  
\begin{equation}
   S_j = 
   \left(
   \begin{array}{cc}
     i r_j & \tau_j\\
     \tau_j & i r_j
   \end{array} 
  \right) 
\:, \ \ \ j=1,2\:,
\label{eq:Sj}
  \end{equation}
where the transmission and reflection coefficients, $\tau_j$ and $r_j$
are assumed to be real. 
In case of $\nu=2$ this scattering matrix relates incoming modes of two outer channels 
(in up/down arms) with the outgoing ones. Electrons in the inner channels are assumed
to propagate through the QPCs without scattering. 

The arms of the interferometer can be different, thereby generally
$x_j^{+}\neq x_j^{-}$.  
The coordinates $L^{\pm}$ and $0$ refer to the points where the MZI is
connected to the drain  
and source reservoirs. Specifying the non-equilibrium boundary
conditions, we assume that only 
one single channel is biased at the chemical potential $eV$, 
while all others are grounded. In case of $\nu=2$ it will be
the outer upper channel, as shown in Fig.~\ref{MZI_scheme}. In the
experiments~\cite{heiblum06,roche07,strunk08} such  
situation is realized with the use of an extra QPC which splits the
incoming inner and outer channels [see Fig.~\ref{MZI_scheme}], so that
these channels originate from different reservoirs.  

To set the stage, we start by considering the simplest situation, when
$r_j=1$ and $\tau_j=0$, so that outer channels in the two 
arms of the interferometer are completely decoupled from each other. 
In this case we model the system by a set of interacting chiral
fermions with the action 
${\cal S} = {\cal S}_0 + {\cal S}_{\rm int}$, where 
\begin{eqnarray}
 {\cal S}_0&=&-i v\sum_{\alpha }\int dt\,dx \,
 \psi^*_{\alpha}(x) \partial_x \psi_{\alpha}(x), \\ 
 {\cal S}_{\rm int}&=& {1\over 2}\sum_{\alpha}\int dt\,dx dx'  \rho_{\alpha} (x) 
U_0 (x-x') \rho_{\alpha}(x') \, . \nonumber
\end{eqnarray}
Here $\psi_{\alpha}$ is a Grassmann field of the
chiral fermion in the arm $\alpha=\pm$. In case of $\nu=2$ it has the
vector structure  
$\psi_{\alpha} = (\psi_{1\alpha}, \psi_{2\alpha})^T$, the subscript
$p=1,2$ denoting the outer and the inner channels, 
respectively. The potential $U_0$ describes the bare interaction
within the interferometer arm between the chiral electron densities 
\begin{equation}
\rho_{\alpha}(x) =\,\psi^*_{\alpha}(x+0)\psi_{\alpha}(x), \nonumber
\end{equation} 
and $v$ denotes the drift velocity in the edge.

In what follows we are going to use the Keldysh version of the
functional bosonization~\cite{lerner04}.  
Let us recall the basic ideas of this construction.
We decouple the interaction term ${\cal S}_{\rm int}$ using the
Hubbard-Stratonovich  
transformation with a field $\varphi$ and double the number of fields; 
$\psi=(\psi_f, \psi_b)^T$ (and the same for $\varphi$), where
$\psi_{f(b)}$ denote the Grassmann fields 
residing at the forward (backward) branch of the Keldysh
contour. These steps lead us to the action 
\begin{eqnarray}
 \mathcal{A}&=&\sum_{\alpha}\left[\int_{{\cal C}_K} dt dx\, \psi^*_\alpha
 (i \partial_t + i v \partial_x - 
\varphi_\alpha)\psi_\alpha \right. \nonumber \\
&+& \left. \frac12 \int_{{\cal C}_K} dt dx dx'\, \varphi_\alpha
  U_0^{-1}(x-x') \varphi_\alpha\right]\:, 
\end{eqnarray}
where we assume the implicit summation over the channel indices.

The minimal coupling between fermionic and bosonic
degrees of freedom in 1D geometry can be eliminated by a local gauge
transformation,  
$\psi_\alpha\rightarrow \psi_\alpha e^{i\Theta_\alpha}$, if one requires that 
\begin{equation}
(\partial_t+v\partial_x) \, \Theta_\alpha=-\varphi_\alpha.
\label{eq:Gauge_eq}
\end{equation}
One has to resolve this differential equation by taking the proper
structure of the Keldysh theory into account, 
\begin{equation}
\left(
\begin{array}{c}
\Theta_f \\
\Theta_b
\end{array}\right)_\xi =
-\int d\xi'\left(
\begin{array}{cc}
D_0^T & D_0^< \\
D_0^> & D_0^{\tilde T}
\end{array}\right)_{\xi-\xi'}
\left(
\begin{array}{c}
\varphi_f \\
-\varphi_b
\end{array}\right)_{\xi'}. \label{eq:Theta}
\end{equation}
Here we have denoted $\xi\equiv(x,t)$ and omitted the arm/channel
indicies for brevity. 
At zero temperature elements of the bare particle-hole propagator $\hat D_0$ 
in the space-time representation are given by the relations
\begin{eqnarray}
D^\gs_{0}(\xi) &=&  v^{-1} \,n_B^\gs(t- x/v), \nonumber \\
D^{T/\tilde T}_{0}(\xi) &=& \theta(\mp t) D^<_{0}(\xi) +\theta(\pm t)
D^>_{0}(\xi), 
\label{eq:DT} 
\label{eq:D0}
\end{eqnarray}
where $n_B^\gs(t)= -i/2\pi(t\mp i a/v)$, $\theta(t)$ is the Heaviside
theta-function  and
$a$ is a short (ultraviolet) cutoff scale that is of the order of magnetic
length $l_B$.  
After this gauge transformation the Green function of interacting
electron acquires the form 
\begin{equation}
 G^>(\xi_1,\xi_2)=g^>(\xi_1-\xi_2) \langle e^{i\Theta_b(\xi_1)} e^{-i
\Theta_f(\xi_2)} \rangle_{\mathcal{A}_0}\,.
\label{eq:G}
\end{equation}
Here $g^{>}(\xi)$ is the free zero-temperature Green function,
\begin{equation}
g^{>}(\xi)=\frac{1}{2\pi v}\frac{ e^{-i(t-x/v) eV} }{( x/v-t+ia/v )}
\label{eq:g0}
\end{equation}
that should be understood as a diagonal $4\times 4$ (two arms, two
channels) matrix; for the sake of generality, we assume that at
$\nu=2$ channels are biased by distinct voltages $V_p^\alpha$, where
$p=1,2$ is the channel index, and $\alpha = \pm$ is the arm index. 
The average over phases $\Theta(\xi)$ in Eq.~(\ref{eq:G}) is performed  
with the Gaussian action
\begin{equation}
 \label{eq:A0}
 \mathcal{A}_0(\varphi)=\frac{1}{2}\vec{\varphi}^{\,t}
 U_0^{-1}\sigma_z \vec{\varphi} - \frac{1}{2} 
\vec{\varphi}^{\,t} \Pi \vec{\varphi} - \vec{\varphi}^{\,t}\sigma_z\rho_0,
\end{equation}
which is a special property of the 1D geometry (Larkin-Dzyaloshinskii
theorem). 
Written in the symbolic form, this expression implies the summation
over the Keldysh  indices (determining the vector structure of $\vec{\varphi}$)
and convolution in the space/time domain, see Eq.~(\ref{eq:Theta}).
The matrix $\Pi$ is the polarization operator ($a,a'$  denoting the
Keldysh indices) 
\begin{equation}
\Pi^{aa'}(\xi) = -i g^{aa'}(\xi)g^{a'a}(-\xi)- (2\pi
v)^{-1}\delta(\xi)\,a\,\delta^{aa'}, 
\end{equation}
with the last term accounting for the static compressibility of edge
channels. The linear in $\varphi$ term 
in the action~(\ref{eq:A0}) describes the response of the system to
the external  
non-equilibrium charge $\vec\rho^{\,t}\equiv (\rho_0,\rho_0)$ injected
from the leads  
\begin{equation}
\rho_0(\xi)=(2\pi v)^{-1}eV(\,t-x/v).
\end{equation}
It is convenient to perform the Keldysh rotation (see e.g.,
Ref.\onlinecite{Kamenev99}), introducing the ``classical" and ``quantum" components  of the fields  
\begin{equation}
\varphi_{c(q)}=(\varphi_f \pm \varphi_b)/2,
\end{equation}
Then the polarization operator in the frequency-momentum representation acquires 
the Keldysh structure with 
\begin{eqnarray}
 &&\Pi_{R,A} = -\frac{\nu}{2\pi} \frac{q}{qv-\omega \mp i0}; \nonumber\\
 &&\Pi_K = (\Pi_R-\Pi_A)\coth(\beta\omega/2). \label{eq:Pi}
\end{eqnarray}
Minimizing the action~(\ref{eq:A0}), 
$\delta \mathcal{A}_0(\varphi)/\delta\varphi_q = 0$, one obtains the
linear equations for  
the mean value of a classical electrostatic field $\bar{\varphi}_c$
and an induced charge $\rho_i$ 
at the edge,
\begin{eqnarray}
 \label{eq:MeanFieldEquation}
  {U}_0^{-1}(q) \bar{\varphi}_c=\rho_i+\rho_0, \nonumber\\
  \rho_i= {\Pi}_R(q,\omega)\bar{\varphi}_c.
\end{eqnarray}
Below we concentrate on the experimental situation with a {\it dc}
applied voltage which introduces a 
homogeneous charge, so that we can take the limit $\omega=0$, $q\sim
1/L$, with $L$ being the system size. 

Evaluation of the Green function~(\ref{eq:G}) is now reduced to a
Gaussian integral:  
\begin{equation}
 G^>(\xi,\xi')=g^>(\xi-\xi')
 \exp\left\{i[\bar{\Theta}(\xi)-\bar{\Theta}(\xi')] -
   J^>(\xi-\xi')\right\}, 
 \label{eq:G_full}
\end{equation}
where the average phase $\bar\Theta$ is induced by the mean classical
electrostatic field 
\begin{equation}
\bar\Theta(\xi)=-v^{-1}\int_0^x dx'\, \bar\varphi_c(x',t+(x'-x)/v),
\label{eq:MeanPhase}
\end{equation}
and $J$ is the correlation function of the Gaussian phase fluctuations
$\delta\Theta$ around the above mean value:
\begin{equation}
J^{aa'}(\xi)=\frac12\langle[\,\delta\Theta^a(\xi)-\delta\Theta^{a'}(0)]^2
\rangle_{\mathcal{A}_0}\,.
\end{equation}

The phase-phase correlation function can be expressed in terms of the
bare particle-hole propagator $D_0$ and the effective interaction
\begin{equation}
\langle \vec\varphi \vec\varphi^{\,t} \rangle = \frac{i}{2}{\cal U} =
\frac{i}{2}(\sigma_z U_0^{-1} - \Pi)^{-1}\,.  
\end{equation}
Using the linear relation~(\ref{eq:Theta}) between the phase and the
electrostatic potential, one obtains
\begin{equation}
J^{aa'}(\xi) = {\cal D}^{aa'}(\xi)-{\cal D}^{aa'}(0),
\end{equation}
where
\begin{equation}
{\cal D} = D_0 {\cal U} D_0. 
\label{eq:D}
\end{equation}
The correlation function $J^{aa'}(\xi)$ can be now easily evaluated
with the use of 
$(\omega,q)$ representation. The result reads
\begin{equation}
 J^{aa'}(\xi)={\nu}^{-1}\left(J^{aa'}_{P}(\xi)-J^{aa'}_{F}(\xi)\right),
 \label{eq:J}
\end{equation}
where we decomposed $J(\xi)$ into plasmon (P) and free (F) parts, 
\begin{equation}
 J^>_{P/F}(x,t)=\int_0^{\infty} \frac{dq}{q}
 \left[1-e^{iqx-iu_{P/F}(q)qt}\right] e^{-a q}, 
 \label{eq:J_pf}
\end{equation}
with the plasmon velocity 
\begin{equation}
u_P(q)=v+\nu U_0(q)/(2\pi)
\end{equation}
and the drift velocity $u_F=v$, respectively. Equation (\ref{eq:J}) is
the zero temperature result for the ``greater" function.
The ``lesser" correlation functions satisfies
$J^<(\xi)=[J^>(\xi)]^*$, and the (anti-) time-ordered correlators can
be reconstructed with the use of basic definitions, akin to
Eq.~(\ref{eq:DT}). 

Equation (\ref{eq:J_pf}) yields for $J_F^>$:
\be
\label{eq:J_f}
J_F^>(\xi) = \ln {x-vt +ia \over ia} \,.
\ee
Thus, in the case $\nu=1$ the bare electron pole in Eq.~(\ref{eq:G}) is
canceled by the free contribution 
to $J(\xi)$ and the electron motion is determined solely by the plasmon
contribution, whereas for $\nu=2$ both plasmon and free
(neutral) modes do influence the behavior of electronic propagator. 
The plasmon contribution $J_P(\xi)$ depends on the specific form
of the electron-electron 
interaction through the momentum dispersion in the velocity $u_P(q)$. 
We consider two models of short-range, $U_0(x)\propto\delta(x)$, and
long-range, $U_0(x)\propto 1/|x|$,  
interactions in the following Sections \ref{s3} and \ref{s4}.

\subsection{Keldysh action of the Mach-Zehnder interferometer}
\label{s2.2}

In this section we generalize the discussion of the functional
bosonization to include electron scattering at QPCs which are crucial
elements of the MZI layout.   
In this way we construct the most general Keldysh action ${\cal A}(\varphi)$ of the MZI, 
which is applicable for arbitrary scattering matrices characterizing two QPCs.
The Keldysh action can be expressed in terms of a total
single-particle time-dependent interferometer 
scattering matrix $S(t,t',[\varphi])$. 
It describes scattering of electrons at QPCs as well as their propagation 
along the arms of the interferometer in the fluctuating
Hubbard-Stratonovich field $\varphi$. 
The $S$-matrix is thus non-local in time and takes different values
($S_f = S[\varphi_f]$ and $S_b = S[\varphi_b]$ ) 
on the forward (backward) branch of the Keldysh contour, since the
fields $\varphi_b(x,t)$ and $\varphi_f(x,t)$ generally differ. The
action ${\cal A}(\varphi)$ has the form  
\begin{eqnarray}
  i\mathcal{A}(\varphi,\vec{\chi})&=&\ln \det
  \left[1+(S_b^{\dagger}e^{i\hat{\chi}}S_f -1)\hat{f} \right] 
  - i\vec\varphi^{\,t}\,\tilde\Pi\,\vec\varphi\, \nonumber\\
  &+& i\vec{\varphi}^{\,t} U_0^{-1}\sigma_x \vec{\varphi}.
  \label{eq:Levitov}
\end{eqnarray}
Here $\vec\varphi = (\varphi_c, \varphi_q)^t$ is assumed to be written in the Keldysh basis.
We have also introduced the electron distribution function in the source
reservoirs, $\hat{f}=\mathrm{diag}(f^{+},f^{-})$ 
(in the case $\nu=2$ each $f^{\pm}$ is itself a diagonal matrix in the
channel space with  the components $f_p^{\pm}$), and auxiliary
{\it ``counting fields''} $\hat{\chi}=\mathrm{diag}(\chi^{+},\chi^{-})$ 
in the drains which enable us to find the number of transferred electrons.
The determinant in Eq.~(\ref{eq:Levitov}) is evaluated with respect to
time, arm, and channel indices. 
Since in this paper we restrict ourselves to the case of zero temperature, 
the distribution functions in time domain read  
\begin{equation}
f_p^{\pm}(t,t')=-\frac{e^{-ieV_p^{\pm}(t-t')}}{2\pi i(t-t'+i0)},
\label{eq:fs}
\end{equation}
where $V_p^{\pm}$ correspond to voltages applied to the different channels. 
The polarization operator $\tilde\Pi$ has the form
\begin{equation}
\tilde\Pi=\left(
\begin{array}{cc}
0 & \Pi_A \\
\Pi_R & 0
\end{array}
\right).
\end{equation}

\begin{figure*}[t]
\centering{
\includegraphics[scale=0.18]{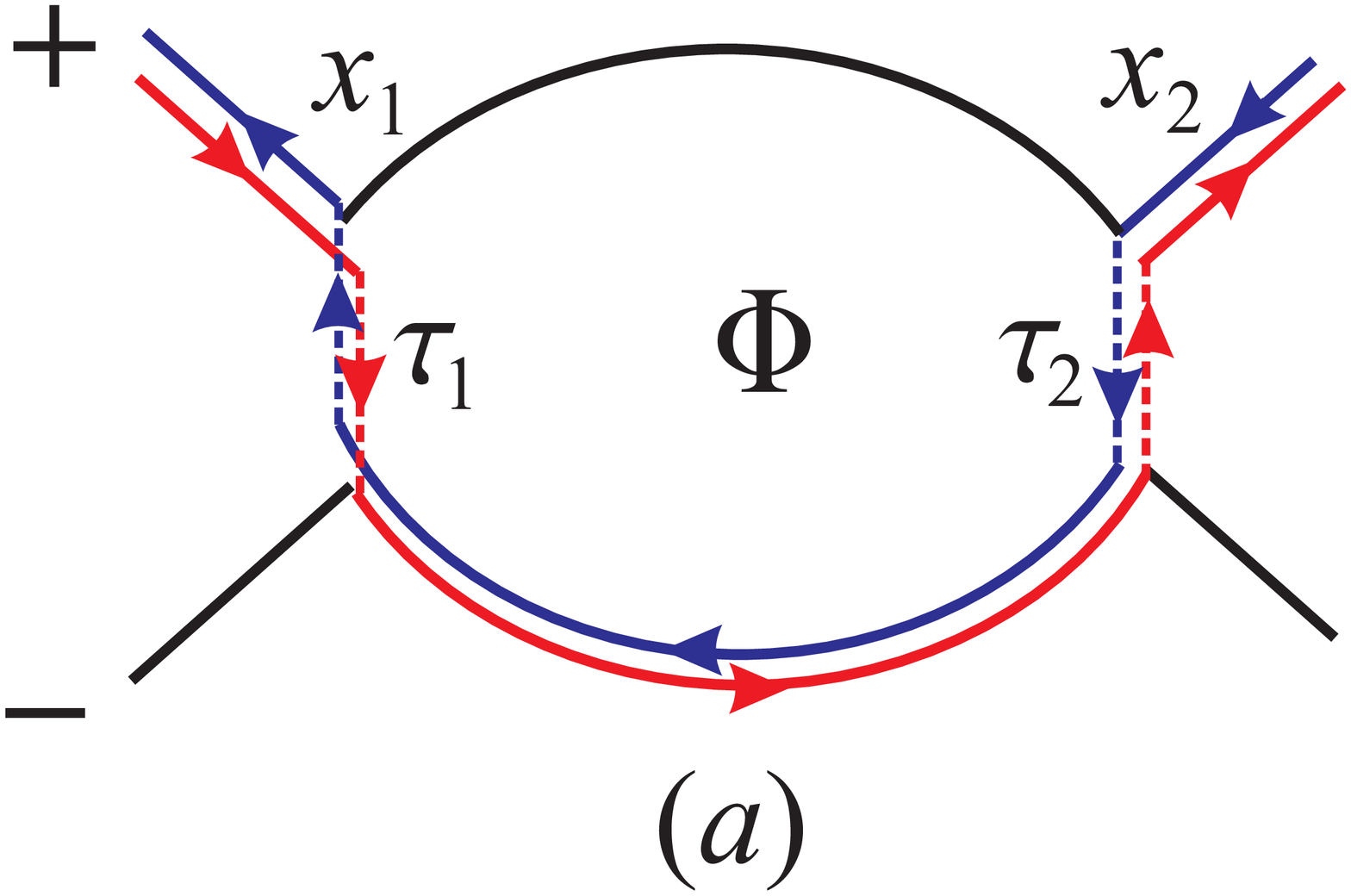}
\hspace{1cm}
\includegraphics[scale=0.18]{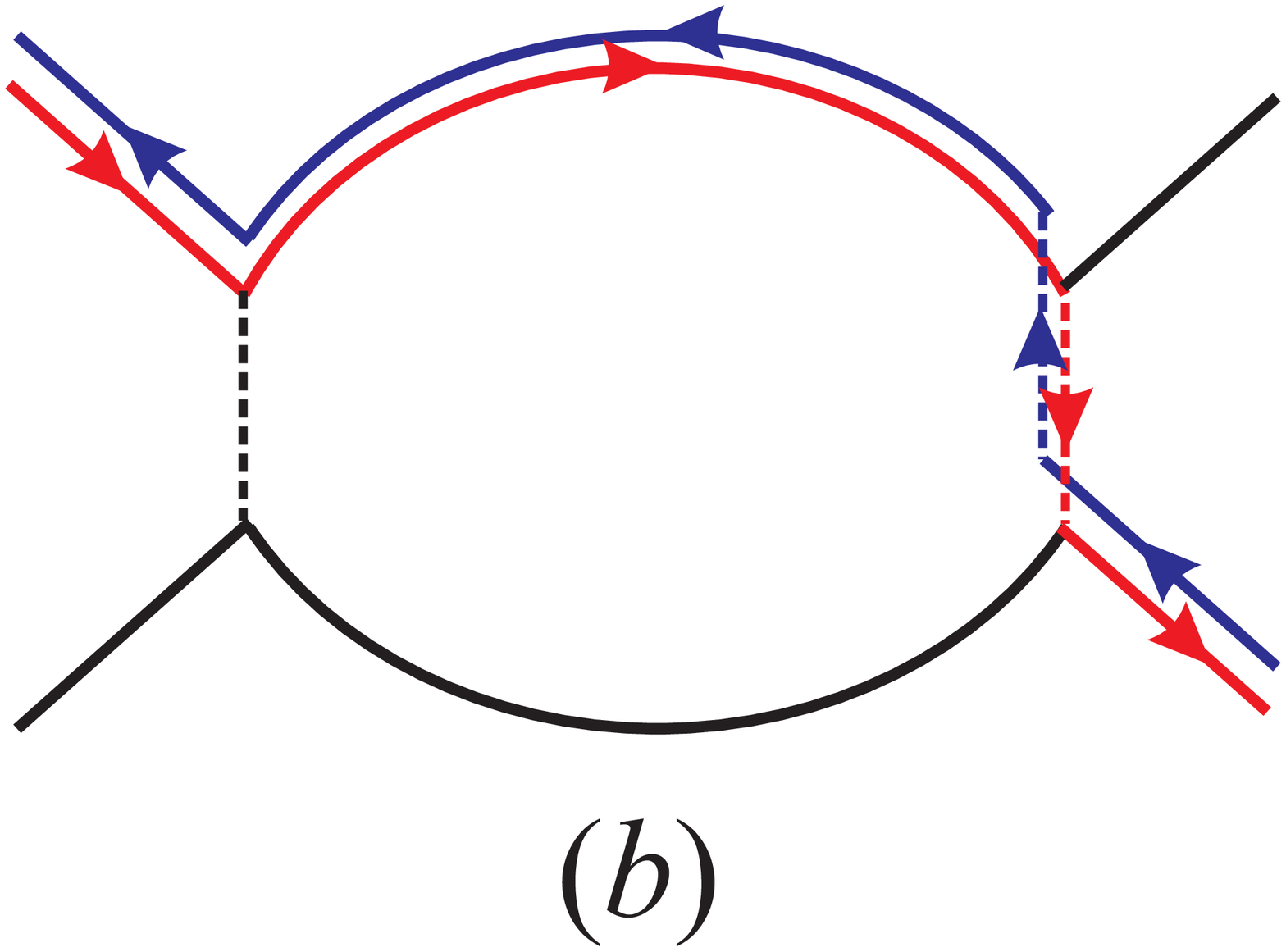}
\hspace{1cm}
\includegraphics[scale=0.18]{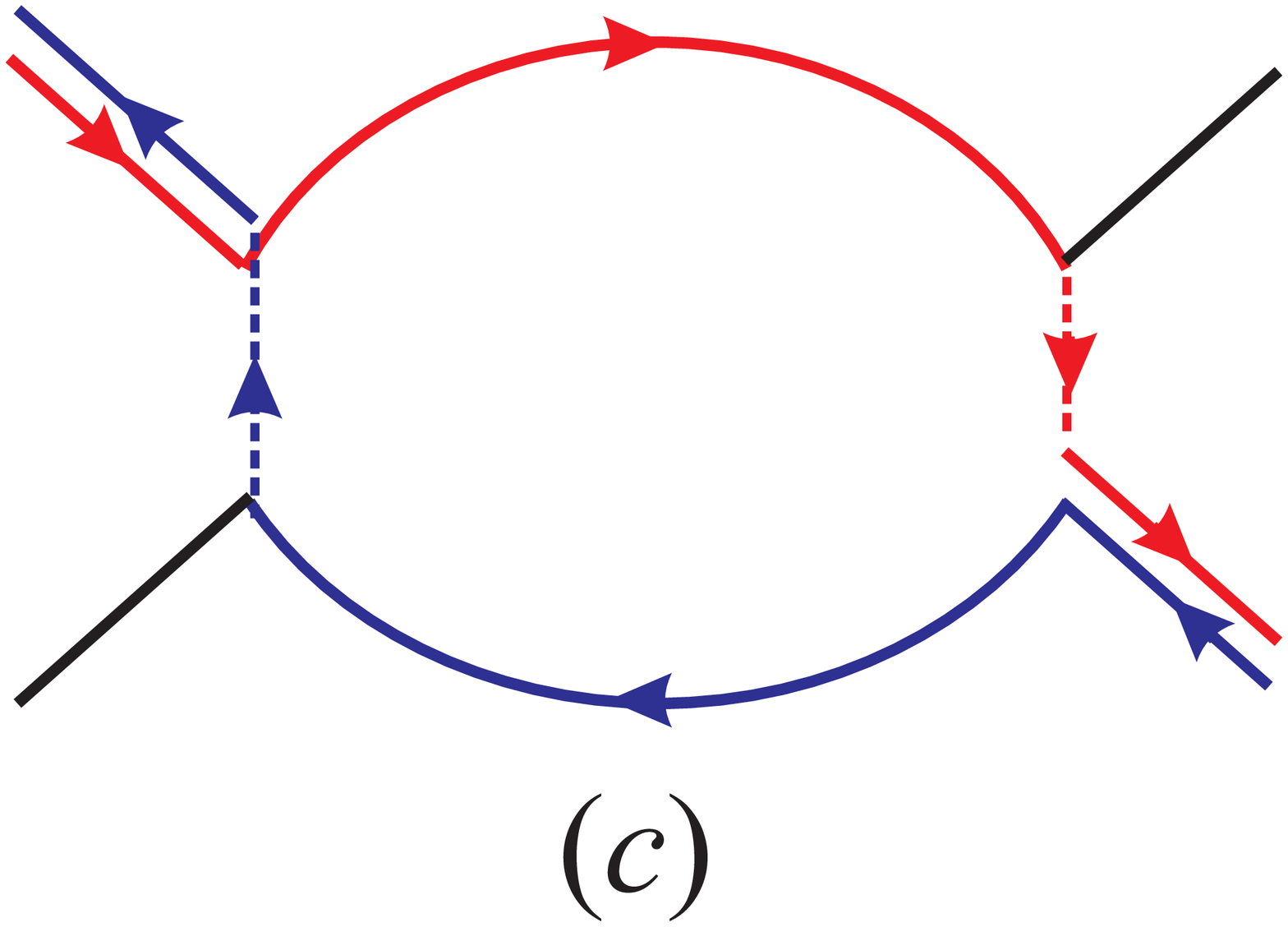}
}
\caption{Example of paths contributing to the matrix element $R=Q_{++}$. 
The paths (a) and (b) give direct contributions, while the path (c)
encloses the magnetic flux $\Phi$ 
and describes the interference process. The corresponding analytical
expressions read: (a)~$\tau_1^2\tau_2^2\Gamma_1^{\dagger
  b}\Gamma_2^b\Gamma_2^{\dagger f}\Gamma_1^f$; 
(b)~$r_1^2\tau_2^2\Gamma_2^{\dagger b}\Gamma_2^{f}e^{-i\chi}$;
(c)~$r_1 r_2 \tau_1 \tau_2 \Gamma_1^{\dagger b} \Gamma_2^{f}e^{i\Phi}
e^{-i\chi}$. } 
\label{fig:Qmn}
\end{figure*}

We now specify the total scattering matrix $S$ for the MZI. It can be
decomposed into the scattering matrices of the quantum point contacts (QPC),
$S_1$ and $S_2$, and the transfer matrices accounting for the propagation 
from $0$ (``source") to $x_1$ ($T_{10}$), from $x_1$ to $x_2$ ($T_{21}$) and from 
$x_2$ to $L$ ($T_{L2}$) (``drain") (see Fig. \ref{fig:MZI}),
\begin{equation}
  S=T_{L2} S_2 T_{21} S_1 T_{10} \, \label{eq:Sfull}.
\end{equation}
with the scattering matrices of the contacts defined by Eq.~(\ref{eq:Sj}).
Since the Hubbard-Stratonovich field is diagonal in the 
channel space, the transfer matrix can be written as 
\begin{equation}
T_{ji}(t,t')=\Lambda_{ji}(t,t')e^{i\theta_{ji}(t')},
\end{equation}
where the shift operator
\begin{equation}
\Lambda^{\pm}_{ji}(t,t')=\delta(t-t'-(x_j^{\pm}-x_i^{\pm})/v)
\end{equation}
describes the free propagation, and 
\begin{equation}
\theta^{\pm}_{ji}(t)=-v^{-1}\int_{x_i^{\pm}}^{x_j^{\pm}} dx'
\varphi^{\pm}(x',t+(x'-x_i^{\pm})/v) \label{eq:Ephase}
\end{equation}
is the electron phase induced by the Hubbard-Stratonovich field during the propagation
of an electron from $x_i$ to $x_j$. 
The weak variation of magnetic field gives rise to an additional
Aharonov-Bohm phase 
$\Phi=\Phi_+ - \Phi_-$, which we incorporate by
$\theta_{21}^{\pm}\rightarrow\theta_{21}^{\pm} + \Phi^{\pm}$. 
Definition~(\ref{eq:Sfull}) gives both $S_f$ and $S_b$ depending on
weather $\varphi_f$ 
or $\varphi_b$ is used to reconstruct the electron phase in
Eq.~(\ref{eq:Ephase}). We note, that the phase $\theta_{f/b}^\pm$, considered
as the function of $(x_j,t)$, satisfies Eq.~(\ref{eq:Gauge_eq}). However,
this phase is different from $\Theta_{f/b}^\pm$, because the relation~(\ref{eq:Ephase})
does not contain information about Keldysh distribution functions, as opposed to
Eq.~(\ref{eq:Theta}).

The cumulant generating function of the MZI can be now expressed as 
a functional integral over $\varphi$,
\begin{equation}
{\cal Z}(\vec\chi) = \int {\cal D}\varphi_{f/b}^{\pm}(x,t)\,
\exp\left\{ i{\cal A}(\,\varphi,\vec\chi) \right\}.
\end{equation}
In particular, it gives the mean number of transferred charges to the
upper/lower drains,    
\begin{equation}
N^{\pm}=-i\partial_{\chi^{\pm}}\ln{\cal Z}(\vec{\chi})|_{\vec{\chi}=0}.
\label{eq:Nmean}
\end{equation}

A detailed derivation of the action~(\ref{eq:Levitov}) will be published
elsewhere~\cite{NgoDinh}. This action bears connection with the solution of
the problem of full counting statistics
\cite{Levitov93,counting-statistics}. Related Keldysh actions of the
determinant structure were found for
the problems of a local scatterer in Ref.~\onlinecite{Snyman08} and of a
quantum wire in Ref.~\onlinecite{gutman10}. 
One can also show that in the limit of zero tunneling, $\tau_i = 0$, 
the full action~(\ref{eq:Levitov}) is reduced to the Gaussian one,
given by Eq.~(\ref{eq:A0}).  
In this case the linear in $\varphi$ term and the classical Gaussian
fluctuations of $\varphi$, described by   
$\Pi_K$ [see Eq.(\ref{eq:Pi})], happen to be the first two terms in
the expansion of the  
functional determinant of the action~(\ref{eq:Levitov}), while the
higher order terms in $\varphi$ vanish.  

In general, the integral over the Hubbard-Stratonovich field with the 
determinant action ${\cal A}(\varphi)$ can
not be evaluated analytically. In the following, 
we proceed with the analysis of the MZI in the weak tunneling limit,
$\tau_j\ll 1$. 
Since in the absence of tunneling the MZI is described by the Gaussian
action~(\ref{eq:A0}), 
we can introduce the tunneling action ${\cal A}_t(\varphi)$, so that
${\cal A}(\varphi) = {\cal A}_0(\varphi) + {\cal A}_t(\varphi)$, where
the expansion of ${\cal A}_t$ in terms of $\tau_j$ starts from the terms 
of order of ${\cal O}(\tau^2)$. In Appendix A we show that  
the tunneling action ${\cal A}_t$ can be obtained by a proper
regularization of the functional determinant appearing in
Eq.~(\ref{eq:Levitov}), yielding
\begin{equation}
 i\mathcal{A}_t=\ln \det \left[ 1+\left(e^{i\tilde{\psi}_b} S_b^{\dagger}(\chi) 
S_f(\chi)e^{-i\tilde{\psi}_f}  -1\right) \bar{f} \right] \, .
\label{eq:At_main}
\end{equation}
Here
\begin{equation}
S_f(\chi)=e^{i\hat{\chi}/2} S_f e^{-i\hat{\chi}/2},\quad
S_b^{\dagger}(\chi)=e^{-i\hat{\chi/2}} S_b^{\dagger} e^{i\hat{\chi/2}},
\end{equation}
and
\begin{eqnarray}
\tilde{\psi}^{\pm}_{f,b}(t)=-v^{-1}\int_{0}^{L^{\pm}} dx'
\varphi^{\pm}_{f,b}(x',t+x'/v) 
\label{eq:psi}
\end{eqnarray}
is the phase collected along the way from $0$ to $L^{\pm}$ without tunneling. 
In other words, $\tilde{\psi}^{\pm}$ is equal to $\theta_{ij}^\pm$ with $x_i=0$ and $x_j=L^\pm$.
New distribution functions $\bar f^{\pm}_\alpha$ are obtained from the
original distribution functions 
in the sources~(\ref{eq:fs}) with the use of a gauge transformation,
\begin{eqnarray}
\bar{f}(t,t')&=&e^{i\hat{\lambda}(t)}\hat{f}(t,t')
e^{-i\hat{\lambda}(t')}, 
\label{eq:f-gauge-transformed}
\\
\hat{\lambda}^\pm_\alpha(t)& = 
&f_>\tilde{\psi}_{f,\alpha}^\pm+f_<\tilde{\psi}_{b,\alpha}^\pm,  
\label{eq:lambda}
\end{eqnarray}
where we defined the projectors on the hole and particle subspaces
\begin{eqnarray}
f_{\gtrless}(t-t')=\pm 1/2\pi i(t-t'\mp i0)
\label{eq:f_gl}
\end{eqnarray}
(see also Appendix A). According to our general notational
conventions, Eq.~(\ref{eq:lambda}) has to be understood as a
convolution in time domain. 

To proceed with the evaluation of the tunneling action, we introduce a
matrix (in the time, arm, and channel space)
\begin{eqnarray}
\label{eq:Q}
Q=e^{i\tilde{\psi}_b} S_b^{\dagger}(\chi) S_f(\chi)e^{-i\tilde{\psi}_f} = 
\left(
\begin{array}{cc}
R & T \\
T' & R'
\end{array}
\right).
\end{eqnarray}
The second equation in Eq.~(\ref{eq:Q}) introduces a block
decomposition of $Q$ in the arm space. 
In the absence of tunneling $Q(t,t')=\delta(t-t')$ and ${\cal A}_t$
vanishes. Since the tunneling affects only outer channels, it is
sufficient to keep only the corresponding elements of the matrix $Q$. 
Therefore, from now on $Q$ is the matrix in the arm and time space only, 
both for $\nu=1$ and $\nu=2$ setups. 

To find $Q$ in the presence of tunneling we define
the ``hopping" operators at each QPC, 
\begin{equation}
\Gamma_{i}^{f/b}(t,t')=e^{i{\theta}_{Li}^{-,f/b}(t)}
\delta(t-t'-(x_i^+-x_i^-)/v)\,
e^{-i{\theta}_{Li}^{+,f/b}(t')},
\label{eq:Gamma}
\end{equation} 
where ${\theta}_{Li}^{\alpha,f/b}(t)$ is the phase picked up by the
electron along the path from $x_i$ to the  
drain $\alpha$, as given by the relation~(\ref{eq:Ephase}).
We can further formulate a set of simple rules that enable one to express
the matrix elements 
$Q_{\alpha\beta}$ in terms of these operators:
\begin{itemize}
\item[(i)]
Each matrix element $Q_{\alpha\beta}$ is associated with a number of
possible continuous paths that originate from  
the source terminal $\alpha$, go to any of the drain terminals and return
back to the source $\beta$  
(see Fig.~\ref{fig:Qmn}). The forward and backward paths are allowed
to go through the 
different interferometer arms. In the latter case the associated
contribution corresponds to the interference process. 
Each element $Q_{\alpha\beta}$ is thus a sum of eight contributions. 
\item[(ii)]
Each transmission via QPC $j$ gives the factor $\tau_j$. Each reflection
gives the factor $\pm i r_j$ along the forward/backward path respectively.
\item[(iii)]
Each tunneling process from arm $+$ to $-$ which happens at the QPC $j$ is
associated with the amplitude $\hat\Gamma_i^{f/b}e^{\mp i \chi/2}$,
where $\chi=\chi_+-\chi_-$. 
The inverse tunneling process is described by the Hermitian conjugate
of this operator. 
The product of such ``hopping" operators must be path-ordered.
\item[(iv)]
contributions associated with the interference process 
carry the extra Aharonov-Bohm phase $e^{\pm i\Phi}$.
\end{itemize}
As an example, three (out of eight) contributions to the matrix
element $R=Q_{++}$  
are shown in Fig.~\ref{fig:Qmn}. The full analytical expressions for
$Q_{\alpha\beta}$ is presented in Appendix B. The next step is to 
use the expansion     
\begin{equation}
\label{eq:Trexpand}
i\mathcal{A}_t = \mathrm{Tr}
\left( (Q-1) \bar{f} -\frac{1}{2}\left((Q-1)\bar{f}\,\right)^2
\right)+ \mathcal{O}(\tau^3) \, . 
\end{equation}
Some important technical details of the evaluation of ${\cal A}_t$
with the use of this formula are  
given in Appendix C. We show there that the second-order
approximation in $\tau_j$ for the tunneling action $\mathcal{A}_t$ has
a form 
\begin{eqnarray}
\label{eq:AES_action}
 i\mathcal{A}_t&=&\sum_{ij}\tau_i\tau_j \\
&\times&\int_{{\cal C}_K} dt_{1,2}\, 
e^{i\Theta(x_i,t_1)}\Pi_{ij}(t_1,t_2)e^{-i\Theta(x_j,t_2)}. \nonumber
\end{eqnarray}
Here the summation is performed with respect to QPC indices ($i,j=1,2$)
and the time integration goes along the Keldysh contour.
The generalized polarization operators for the direct terms read
\begin{eqnarray}
 \label{eq:Pol1}
 \Pi_{ii}^{T / \tilde T}(t_1, t_2) &=&  
 - \, f^{+}_{>,<}(t_1-t_2)f^{-}_{<,>}(t_2-t_1)\,, \nonumber \\
 \Pi_{ii}^{\gtrless}(t_1, t_2) &=&  \Pi_{ii}^{T/\tilde{T}}(t_1, t_2)\, e^{\pm
i\chi},  \quad
\end{eqnarray}
and the polarization operators of the interference terms are given by
\begin{eqnarray}
 \label{eq:Pol2}
 \Pi_{12}^{T / \tilde T}(t_1, t_2) &=&  
 -  e^{-i\Phi} f^{+}_{<,>}\left(t_1-t_2-(x_1^+-x_2^+)/v\right)  \nonumber \\  
 & &\times f^{-}_{>,<}\left(t_2-t_1-(x_2^- -x_1^-)/v\right)\,, \nonumber \\
 \Pi_{12}^{\gtrless}(t_1, t_2) &=&  \Pi_{12}^{\tilde{T}/T}(t_1, t_2)\, e^{\pm
i\chi}, \\  
 \label{eq:Pol3}
 \Pi_{21}^{T / \tilde T}(t_1, t_2) &=&  -e^{i\Phi}
 f^{+}_{>,<}\left(t_1-t_2-(x_2^+-x_1^+)/v\right) \nonumber\\
 & &\times f^{-}_{<,>}\left(t_2-t_1-(x_1^- -x_2^-)/v\right)\,, \nonumber\\
 \Pi_{21}^{\gtrless}(t_1, t_2) &=&  \Pi_{21}^{T/\tilde{T}}(t_1, t_2)\, e^{\pm
i\chi}.  \quad
\end{eqnarray} 
They are built up from the electron and hole 
distribution functions
\begin{equation}
f_{\gtrless}^{\pm}(t-t')= e^{-ieV^{\pm}(t-t')} f_{\gtrless}(t-t').
\end{equation}
The polarization operators are dressed by relative gauge factors
$\Theta(x_i,t)=\Theta^{+}(x_i^+,t)-\Theta^{-}(x_i^-,t)$, where the
phases $\Theta^\pm$  
at each arms $\pm$ are linked to the corresponding HS-field $\varphi^\pm$ 
by relation~(\ref{eq:Theta}), thereby taking the interaction into account.

The structure of the polarization operators can be interpreted as follows. 
The component $\Pi^{<}\propto f^+_< f^-_>$ corresponds to a particle
excitation in the arm $+$ 
and a hole excitation in the arm $-$. This implies the charge transfer
from $+$ to $-$, which explains the ``counting" factor
$e^{-i\chi}$. Similar considerations  explain the structure of $\Pi^{>}$.
Finally, the components $\Pi^{T/\tilde{T}}$ at $\chi=0$ are related to
$\Pi^{\gtrless}$ via the conventional Keldysh-formalism relations, 
$\Pi^{T/\tilde{T}}\bigl|_{\chi=0}(t)=\theta(t)\Pi^{\gtrless}(t)+\theta(-t)\Pi^{\lessgtr}(t)$. 

\subsection{Gaussian approximation}
\label{s2.3}

The Keldysh action enables us to calculate the tunneling current of the
interferometer. Specifically, let us consider the
current of particles transported from the upper ($+$) to the lower
interferometer arm ($-$). This current can be obtained as derivative of the 
cumulant generating function ${\cal Z}(\vec\chi)$  
with respect to $\chi_-$. We express the current as difference of up-
and down rates, 
$I=\sum_{ij}(I_{ij}^{<}-I_{ij}^{>})$, with
\begin{eqnarray}
\label{eq:current}
 I_{ij}^{<,>}=-\frac{e}{t_0}\tau_i\tau_j\int_{-\infty}^{\infty} dt_1 dt_2
\Pi_{ij}^{<,>}(t_1,t_2) \nonumber\\ 
\times \left\langle e^{i\Theta^{f,b}(x_i,t_1)} e^{-i\Theta^{b,f}(x_j,t_2)}
\right \rangle_{\mathcal{A}_0+\mathcal{A}_t(\chi=0)}\,,
\end{eqnarray}
where we introduced the measurement time $t_0$.
The average over phases is to be done with the full action. However,
to the leading
approximation, we can neglect the tunneling action and perform the
phase average with the Gaussian action ${\cal A}_0$. The corresponding
correlation  functions of phase variables are given by
Eq.~(\ref{eq:J}). Assuming that no interaction is present between the 
different interferometer arms, we find
\begin{eqnarray}
\label{eq:cuurent0}
 I_{ij}^{\gtrless}&=&e \tau_i\tau_j v^2  \label{eq:current1} \\ 
&\times&\int dt
G^{+,\gtrless}(x_i^+-x_j^+,t) G^{-,\lessgtr}(x_j^--x_i^-,-t)\,, \nonumber 
\end{eqnarray}
where the full Green functions are given by Eq.~(\ref{eq:G_full}).
Equation (\ref{eq:cuurent0}) 
was the starting point of the discussion in the works by
Chalker, Gefen and Veillette~\cite{Chalker07} and 
Levkivskyi and Sukhorukov~\cite{Levkivskyi08}. We go beyond it by
including into consideration the (non-Gaussian) tunneling action that
controls the non-equilibrium dephasing.

\subsection{Non-Gaussian effects}
\label{s2.4}

Let us now improve the result for the tunneling current~(\ref{eq:current1})
by taking into account the non-Gaussian nature of phase fluctuations,
which are described by the tunneling part of the action ${\cal A}_t$. These
fluctuations are responsible for the intrinsic dephasing in the MZI by
the non-equilibrium shot noise generated in the QPCs.
To be specific, we evaluate $I_{ij}^<$, which requires calculation of
the correlation function
\begin{equation}
 \left\langle e^{i\Theta^{f}(x_i,t_1)} e^{-i\Theta^{b}(x_j,t_2)} \right
\rangle=\int\mathcal{D}\varphi
e^{i\mathcal{A}^<_{ij}+i\mathcal{A}_0+i\mathcal{A}_{t,\chi=0}},
\label{eq:CorrPathInt}
\end{equation}
where we have introduced a linear-in-$\varphi$ term in the action,
\begin{eqnarray}
\mathcal{A}^<_{ij}(t_1,t_2,[\varphi]) &=&  \Theta^{f}(x_i,t_1) -
\Theta^{b}(x_j,t_2) \nonumber \\ 
&=& \sum_{\alpha} \int dtdx (\vec\Theta^{\alpha})^t \vec{\cal J}^\alpha.
\end{eqnarray}
Here $\vec{\cal J}^\pm = ({\cal J}^{\pm,f}, {\cal J}^{\pm,b})^t$  are source terms acting on 
both branches of the Keldysh contour, 
\begin{eqnarray}
 \mathcal{J}^{\pm,f}(x,t)&=\pm\delta(x-x^{\pm}_{i})\delta(t-t_1),\nonumber\\
 \mathcal{J}^{\pm,b}(x,t)&=\mp\delta(x-x^{\pm}_{j})\delta(t-t_2).
\end{eqnarray}

In order to account for the non-Gaussian effects, we proceed with a
real-time instanton approach developed in Ref.~\onlinecite{NgoDinh10} for
the problem of tunneling spectroscopy  of a biased quantum wire with a
scatterer. 
We look for the optimal (saddle-point) trajectory $\varphi^\alpha_*(x,t)$, which
minimizes the total action
$\mathcal{A}_{\rm
  tot}=\mathcal{A}_0+\mathcal{A}_{t}+\mathcal{A}_{ij}^<$ at $\chi=0$. 
In the weak tunneling limit, $\tau_j\ll 1$, one can find the instanton
trajectory $\varphi^*$ 
approximately by minimizing only the quadratic part of the action, i.e. 
$\mathcal{A}_0 + \mathcal{A}_{ij}^<$.
This yields for the stationary point 
\begin{equation}
\vec\Theta^\alpha[\varphi^\alpha_*]=-{\cal D}\vec{\mathcal J}^\alpha, 
\end{equation}
where the particle-hole propagator is defined in Eq.~(\ref{eq:D}).
One can show that taking into account
corrections of order $\tau^2$ which follow from the exact non-linear equation
of motion would lead to a contribution of order $\tau^4$ to the tunneling
action. Thus, such corrections are negligible within the accuracy
$\tau^2$ of our calculation. 

For simplicity, we do not take into account at this stage charging
effects on the interferometer arms.  
These effects will be restored at the end of the calculation as 
constant-in-time phase shifts evaluated according to
Eq.~(\ref{eq:MeanPhase}).

\begin{figure}[b]
\includegraphics[width=2.3in]{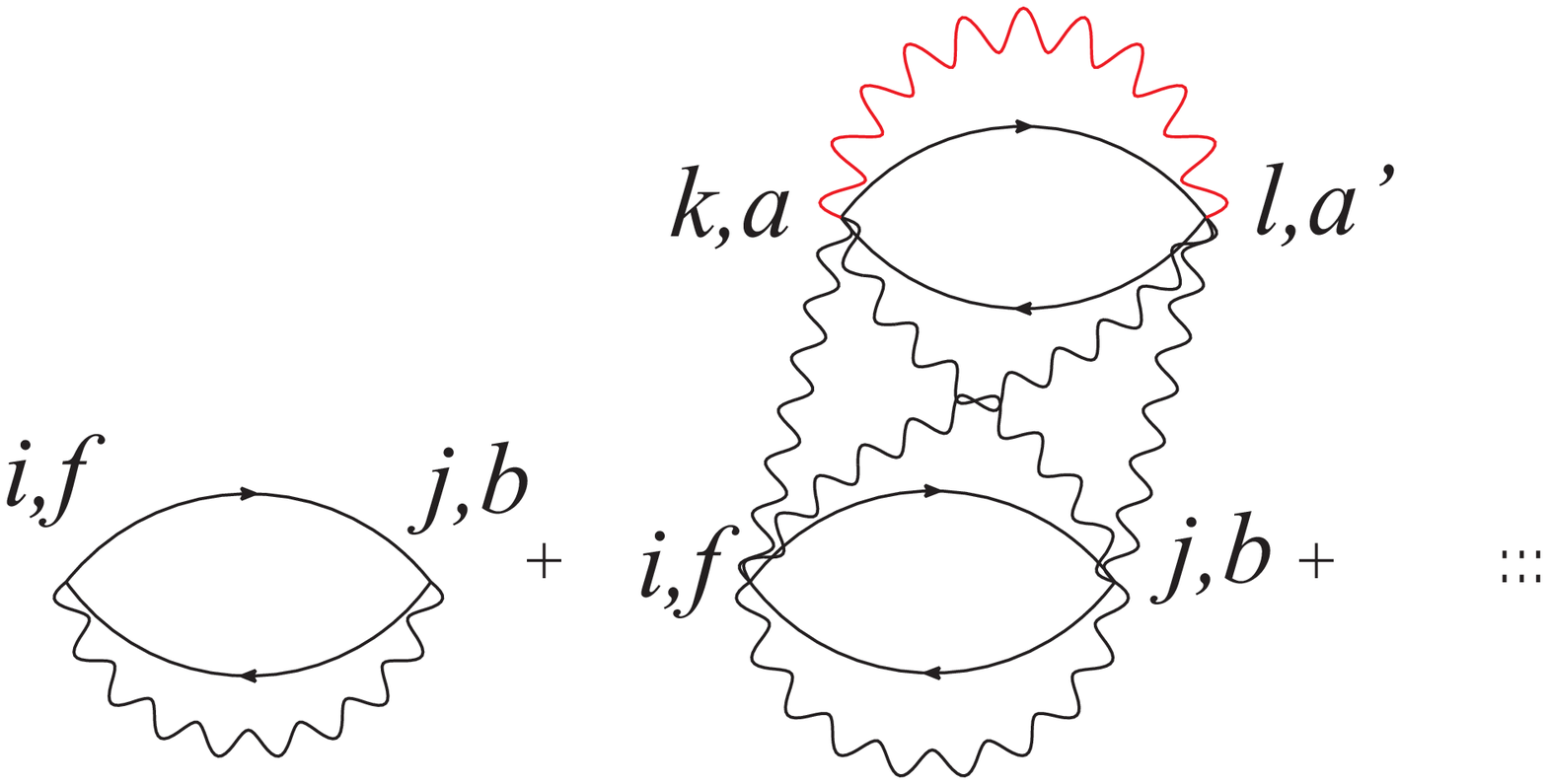}
\caption{Diagrammatic series for the tunneling current corresponding
  to the real-time instanton  
configuration~(\ref{eq:Phi_Inst}). 
Black wavy lines, representing the exponent of
the phase correlation 
function, $\exp\{-J(\xi)\}$, 
are taken into account by the saddle-point approximation, while the
red ones originate from the quantum fluctuations around the instanton.}
\label{fig:Diagrams}
\end{figure}

Following this procedure, we find for the tunneling action in the saddle-point
approximation
\begin{eqnarray}
\label{eq:TW_SP}
 i\mathcal{A}^{*,<}_{t,ij}&=&\sum_{kl}\tau_k\tau_l 
 \int_{{\cal C}_K} dt'_{1,2} 
 e^{i\Theta_k^{*}(t_1')}\tilde\Pi_{kl}(t'_1,t'_2)e^{-i\Theta_l^{*}(t'_2)}. \nonumber
 \\ 
\end{eqnarray}
with the phases
\begin{eqnarray}
i\Theta_k^{*,a}(t)&=& J^{fa} (x_{i}^+ -x_{k}^+,t_1-t) -  J^{ba}
(x_{j}^+ -x_{k}^+,t_2-t) \nonumber\\
   &+&  J^{fa} (x_{i}^- -x_{k}^-,t_1-t) -  J^{ba } (x_{j}^-
-x_{k}^-,t_2-t)\nonumber\\
\label{eq:Phi_Inst}
\end{eqnarray}
($a$ is the Keldysh index).
In the stationary situation this action should become a function of
time difference only,  
$\mathcal{A}^{*,<}_{t,ij}(t_1-t_2)$.
We have also taken into account quantum fluctuations around the
instanton trajectory  
$\varphi^\alpha_*$. One can prove that, in analogy with
Ref.~\onlinecite{NgoDinh10},  they result in renormalization of the  
bare polarization operator $\Pi_{kl}$ to the dressed one given by
\begin{eqnarray}
\tilde{\Pi}^{aa'}_{kl}(t_1,t_2) &=& {\Pi}^{aa'}_{kl}(t_1,t_2)\nonumber\\
&\times & \exp\{ -J^{aa'}(t_1-t_2, x^+_k-x^-_l)\}\nonumber\\
&\times & \exp\{ -J^{a'a}(t_2-t_1, x^-_l-x^-_k)\}.
\end{eqnarray}

The tunneling current can be now evaluated by substituting the stationary
phase $\Theta^*$ into relations~(\ref{eq:current}) and (\ref{eq:CorrPathInt}).
The diagrammatic representation resulting from expansion of the
exponential of the tunneling action evaluated in the saddle-point
approximation is shown in Fig.~\ref{fig:Diagrams}. 
If we retain only the first term of this expansion (i.e. neglect the
tunneling action), we  reproduce
the result~(\ref{eq:current1}) of the previous section. The non-Gaussian
shot noise modifies this result by the additional factor
$e^{i\mathcal{A}_t^*}$ under the time integral,
\begin{eqnarray}
 I_{ij}^{\gtrless}&=&e \tau_i\tau_j v^2 \int dt' \exp\{iA_{t,ij}^{*,\gtrless}(t')\}
 \label{eq:current2} \\ 
&\times&
G^{+,\gtrless}(x_i^+-x_j^+,t') G^{-,\lessgtr}(x_j^--x_i^-,-t'). \nonumber 
\end{eqnarray}

The full Green's function $G^+$ here contains the extra phase shift,
which takes into account the non-equilibrium charge effects.  It has to be found in accordance 
with Eqs.~(\ref{eq:MeanFieldEquation}) and (\ref{eq:MeanPhase}). Solving mean field 
equations~(\ref{eq:MeanFieldEquation}) in the {\it dc} limit, i.e. at $\omega=0$, $q\sim 1/L$, 
with the use of $\rho_0 = eV/(2\pi v)$ as the injected non-equilibrium charge, one gets
the mean electrostatic potential on the upper arm of the interferometer,
\begin{equation}
 \bar{\varphi}_c^+=\nu^{-1}\left(1-\frac{v}{u_{P}(1/L)}\right) eV
\end{equation}
and the non-equilibrium phase shift
\begin{equation}
 \bar{\Theta}_c^+(x)=\nu^{-1}\left(\frac{1}{u_{P}(1/L)} - \frac{1}{v}\right) eV x.
 \label{eq:MTheta}
\end{equation}

Using the expression~(\ref{eq:current2}), we can now derive a general result for the
differential conductance ${\cal G}(V)=dI/dV$. Specifically, 
the conductance is obtained as a sum of two contributions: ${\cal G}_0(V)$ which
is independent on the enclosed flux $\Phi$ and another one, ${\cal G}_\Phi(V)$, 
which is sensitive to $\Phi$,
\begin{equation}
\mathcal{G}(V)=\frac{e^2}{2\pi\hbar}\Bigl( \tau_1^2 \mathcal{I}_1
+\tau_2^2 \mathcal{I}_2 +  
 2 \tau_1 \tau_2 {\rm Re}\left( \mathcal{I}_{\Phi} e^{i\Phi}\right) \Bigr)\,,
\label{eq:GV}
\end{equation}
where
\begin{eqnarray}
 \mathcal{I}_k &\simeq&  4\pi v^2 \int dt \, t e^{-ieVt} \mathrm{Im}
 (G_0^{>}(0,t))^2 
 \exp\left\{ - {\rm Im} \, A_{t, kk}^{*}(t)\right\}\,,  \nonumber \\
 \mathcal{I}_{\Phi} &\simeq& 4\pi v^2  \int dt (t-\Delta t)\,
 e^{-ieV(t-\Delta t)} \exp\{ - {\rm Im} \, A_{t,12}^{*}(t)\}
 \nonumber\\ 
&\times&  \mathrm{Im} (G_0^{>}(l^+,t) G_0^{<}(-l^-,-t) ). 
\label{eq:I}
\end{eqnarray}
In this expression $l^\pm = x_2^\pm - x_1^\pm$ are the lengths of the
upper and lower arms of the MZI, and 
\begin{equation}
 \Delta t = \frac{\partial}{\partial (eV)} \Bigl[\,
 \bar\Theta^+_c(x_2^+)-\bar\Theta^+_c(x_1^+)\Bigr] 
 + {l^+}/{v}
\label{eq:Dt}
\end{equation}
is the delay time related to the charging effects discussed in 
Sec.~\ref{s2.1}. The explicit form of the delay time $\Delta t$
depends on specific model of the interaction. When deriving 
the result~(\ref{eq:I}), we have neglected the real part of the
tunneling action, ${\rm Re}\, A_t^*$ (it leads to small corrections only
that do not affect the result in any essential way).

\section{Short-range interaction}
\label{s3}

In this section we analyze  a model  of an interferometer with
$\nu=2$ channels per 
interferometer arm and short-range (point-like) interaction, $U_0(x-x')=U_0
\delta(x-x')$. 
This model was considered previously in Ref.~\onlinecite{Levkivskyi08} where
non-equilibrium dephasing effects were discarded. 
We will show that  predictions of this model can not be reconciled 
with the experiments, which can be traced back to the simplified
treatment of the interaction. Nevertheless, it is natural to start our
analysis from this model that serves nicely for illustration of
general principles. 

The elementary excitations induced by the short-range interaction
interaction are  
dispersiveness collective charge and neutral modes with velocities
$u=v_F+{U_0}/{\pi}$ and $v$ respectively. 
The correlation functions read
\begin{equation}
 J^{\gtrless}(x,t)=\frac{1}{2}\ln\left(\frac{x-u t_{\mp}}{x-v t_{\mp}}\right),
 \label{eq:J_g_sr}
\end{equation}
\begin{equation}
 J^{T/\tilde{T}}(x,t)=\theta(t)J^{\gtrless}(x,t)+\theta(-t)J^{\lessgtr}(x,t),
\end{equation}
where $t_{\pm}=t\pm i0$.
It is easy to verify that Eq.~(\ref{eq:J_g_sr}) can be also written as
\begin{equation}
 \label{eq:SymmetryPropertiesJ}
 J^{T/\tilde{T}}(x,t)=\theta(x)J^{\gtrless}(x,t)+\theta(-x)J^{\lessgtr}(x,t)\,.
\end{equation}
The last equation is a consequence of the chirality
of the system. Similar relations hold also for the polarization
operators $\Pi_{kl}$, and 
these relations remain unchanged after renormalization. 

Using the relations~(\ref{eq:SymmetryPropertiesJ})
for the correlation functions, 
one can show that the saddle-point phase $\Theta^{*,a}_k$ does not
depend on the Keldysh  
index $a$ if $k=2$. (We remind that, according to our convention, $x_2^\pm
> x_1^\pm$.)
Analogously, the renormalized 
polarization operator $\tilde{\Pi}^{aa'}_{kl}$ does not depend on $a$
($a'$), if $k=2$ ($l=2$).  
Using the above relations and taking into account contributions from
the forward and backward Keldysh contour,  
we find that  the only non-vanishing term in the sum~(\ref{eq:TW_SP})
for the tunneling action $\mathcal{A}_t^*$
is the one with $k=l=1$. In other words, the tunneling action in 
the saddle-point approximation has the only contribution $\propto
\tau_1^2$, which means that  
only the noise produced by the first contact contributes to
dephasing. This remarkable property has a 
simple physical explanation: in chiral 1D-system particles possess a well-defined
prehistory, where the direction in space is linked to the direction of time. 
At zero temperature but finite voltage the tunneling of electrons via
the first QPC is accompanied by a spontaneous emission of non-equilibrium
plasmons and neutral modes. 
They transfer the shot noise $\propto \tau_1^2$ to the second QPC, where 
it affects the direct ($I_{22}$) and interference ($I_{12}$)
current. On the other hand, the shot noise generated  
at the second QPC is transferred directly to the drains without
affecting the particle interference. 

Let us evaluate the tunneling action for the interference current $I_{12}^{<}$,
\begin{equation}
 \label{eq:Tunnelingaction}
  i \mathcal{A}_{t,12}^{*} =  \tau_{1}^2 \int dt'_{1,2} 
   e^{i\Theta_1^{*}(t'_1)}
    \tilde{\Pi}_{11} (t'_1-t'_2) e^{-i\Theta_1^{*}(t'_2)},
\end{equation}
\begin{equation}
e^{i\Theta_1^{*f/b}(t)}=\frac{(t+\frac{l^+}{v}\pm
i0)^{\frac{1}{2}}(t+\frac{l^-}{v}\pm i0)^{\frac{1}{2}}}{(t+\frac{l^+}{u}\pm
i0)^{\frac{1}{2}}(t+\frac{l^-}{u}\pm i0)^{\frac{1}{2}}}, 
\end{equation}
where $l^{\pm}=x_2^{\pm}-x_1^{\pm}$ are the lengths of the interferometer arms.
Remarkably, this action for the interference current does not depend on the
times $t_{1,2}$, and therefore, the current factorizes in the one
found in Sec.~\ref{s2.3} and a dephasing factor.
The dominant large-voltage behavior of (\ref{eq:Tunnelingaction}) is determined
by the singularities of the polarization operator at $t'_1\sim t'_2 \sim 1/eV$, yielding
\begin{eqnarray}
  i \mathcal{A}_{t,12}^{*} =  -\tau_{1}^2 \int_{-\infty}^{\infty} dt
[\tilde{P}_{11}^{>} (e^{i\Theta_1^{*b}(t)-i\Theta_1^{*f}(t)}-1)\nonumber\\
+\tilde{P}_{11}^{<} (e^{i\Theta_1^{*f}(t)-i\Theta_1^{*b}(t)}-1)],
\end{eqnarray}
where 
\begin{eqnarray}
\tilde{P}_{11}^{\gtrless}=\int dt
\tilde{\Pi}_{11}^{\gtrless}(t)=-\frac{v}{u}\theta(\mp eV)
\frac{|eV|}{2\pi}.
\end{eqnarray}
Thus, for the imaginary part of the action we have 
\begin{eqnarray}
\label{eq:Im_At_12}
 \mathrm{Im}\, \mathcal{A}_{t,12}^{*} = &-& 2\tau_{1}^2 
[\tilde{P}_{11}^{>}+\tilde{P}_{11}^{<}] \\
 &\times& \int_{-\infty}^{\infty} dt
\sin^2\left(\frac{\Theta_1^{*b}(t)-\Theta_1^{*f}(t)}{2}\right). \nonumber
\end{eqnarray}
The only effect of quantum corrections to the
saddle-point approximation is a renormalization of the tunneling strength,
$\tau_1^2\rightarrow\tilde{\tau}_1^2=\tau_1^2 v/u$. 

The final result for the dephasing rate 
depends on relation between the parameters of the problem.
In the case  $l^+/l^->u/v$, we find
\begin{equation}
 \mathrm{Im}\, \mathcal{A}_{t,12}^{*}=
 \tilde{\tau}_1^2\frac{1}{\pi}|eV|(l^+ +l^-) 
\left(\frac{1}{v}-\frac{1}{u}\right).
\end{equation}
Therefore, in this regime the dephasing rate is proportional to 
the total length of the
interferometer. Clearly, there is no dephasing when the velocities are
equal to each other,
which is the case for the non-interacting system.
If parameters are chosen in such a way that $l^+/l^-<u/v$, we
get
\begin{equation}
 \mathrm{Im} \mathcal{A}_{t,12}^{*}= \tilde{\tau}_1^2\frac{1}{\pi}|eV||l^+ -l^-|
\left(\frac{1}{v}+\frac{1}{u}\right).
\label{eq:At_12}
\end{equation}
Remarkably, we find no dephasing in the case of equal arm
lengths. We offer the following physical interpretation of this
result. A tunneling event at the
first contact is accompanied by emission of plasmons in the upper and lower
interferometer arms. If the arm lengths are equal, 
the plasmons meet together at
the second contact completely phase-coherently. In other words, when electrons
tunnel at the first contact, there will be noise created in the upper and lower
interferometer arm, and this noise is fully correlated. In the case of equal arm
lengths it is possible to restore the complete phase information at the second
contact. Thus, dephasing is absent.

\begin{figure}[t]
\includegraphics[width=3.0in]{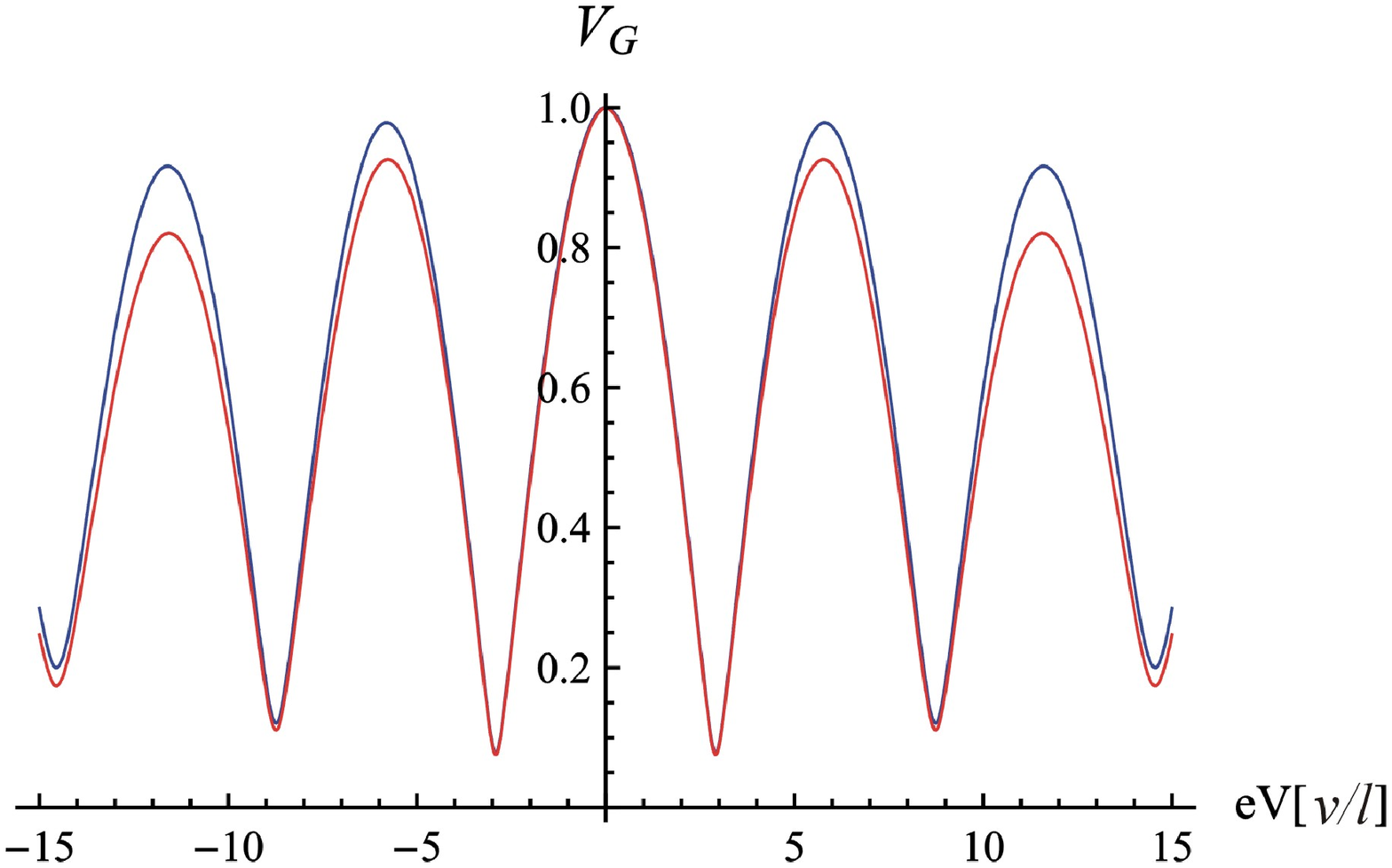}
\includegraphics[width=3.0in]{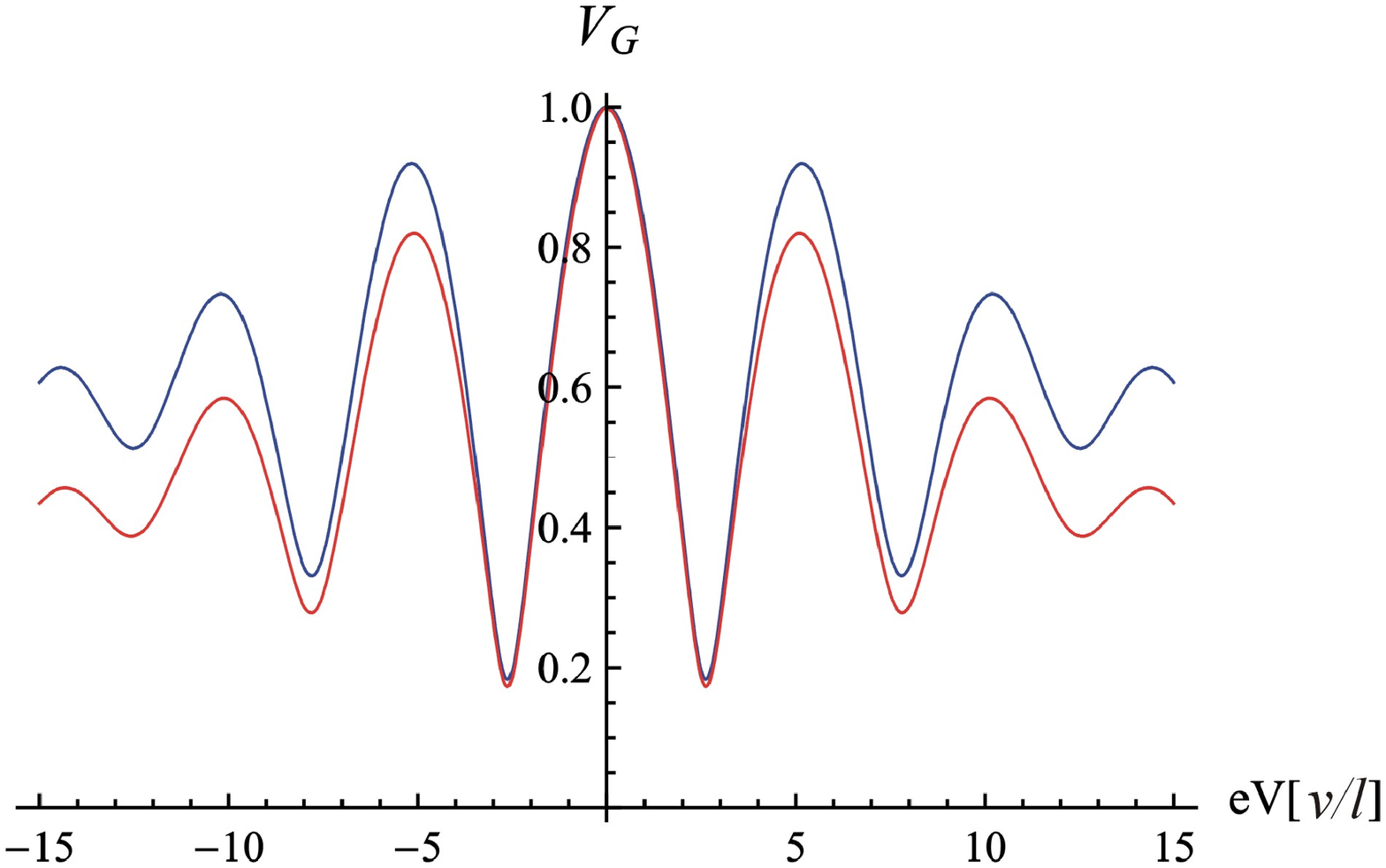}
\caption{Visibility at $\nu=2$ calculated in the model of strong short-range interaction, $u\gg v$,  
with equal QPCs, $\tilde\tau_1^2 = 0.2$. Blue curve: dephasing is
neglected; red curve: dephasing by the shot-noise  
is taken into account. Upper panel: $l^+/l^- = 1.15$; lower panel:
$l^+/l^- = 1.35$.} 
\label{fig:Vis_SR}
\end{figure}

In the case of direct currents, an analogous calculation for the
long-time limit gives 
\begin{eqnarray}
 \mathrm{Im}\, \mathcal{A}_{t,11}^{*}(t) &=& 0, \\
 \mathrm{Im}\, \mathcal{A}_{t,22}^{*}(t) &=&
 \tilde{\tau}_1^2\frac{4}{\pi}|eV t|. \nonumber 
\end{eqnarray}
As for the time delay $\Delta t$ related to the charging of the upper
arm of the interferometer, we obtain, 
using Eqs.~(\ref{eq:MeanFieldEquation}), (\ref{eq:MeanPhase}) and
(\ref{eq:Dt}),
\begin{equation}
 \Delta t = \frac{l^+}{2}\left(\frac1u + \frac 1v \right).
\end{equation}

We now have all the ingredients needed to evaluate the visibility of the MZI
according to Eqs.~(\ref{eq:GV}) 
and (\ref{eq:I}) for the differential conductance. We define the
voltage dependent visibility as the ratio of the amplitude of Aharonov-Bohm oscillations in conductance
to its mean value:
\begin{equation}
 {\cal V}_G(V) = \Bigl( \max({\cal G}_\Phi(V)) -\min({\cal G}_\Phi(V))\Bigr)
 /2\,{\cal G}_0(V). 
 \label{eq:Vis_def}
\end{equation}
We observe that the direct conductance ${\cal G}_0(V)$
does not depend on voltage 
despite a non-zero action $A^*_{t,22}$: the latter affects each of the
tunneling rates $I^>_{22}$ and $I^<_{22}$ 
but does not change their difference.  Therefore, the visibility can
be expressed via  
the integral ${\cal I}_\Phi$ given by the relation~(\ref{eq:I}) as
\begin{equation}
 {\cal V}_G(V) = \frac{2\tau_1\tau_2}{\tau_1^2 + \tau_2^2} |{\cal I}_\Phi|\,.
\end{equation}
In what follows we consider the limit of strong interaction, $u\gg
v$. In this case we represent ${\cal I}_\Phi$ 
in the form 
\begin{equation}
 {\cal I}_\Phi  = {\cal I}_\Phi^{(0)}\exp\{- {\rm Im}\, {\cal A}^*_{t,12}\},
\end{equation}
where ${\cal I}_\Phi^{(0)}$ is simplified to
\begin{eqnarray}
 \label{eq:IntI}
 \mathcal{I}_\Phi^{(0)} = &-& \int \frac{dt}{\pi} (t-\Delta
 t)e^{-ieV(t-\Delta t) } \\ 
&\times& {\rm Im} \left[
\frac{1}{(t+i0)(t-\frac{\Delta x^+}{v}
  +i0)^{\frac{1}{2}}(t-\frac{\Delta x^-}{v} +i0)^{\frac{1}{2}}} 
\right]\,. \nonumber
\end{eqnarray}
Two examples of the visibility plots in the strong-interaction limit
and equal transmissions at both QPCs are shown 
in Fig.~\ref{fig:Vis_SR}, where we have denoted $l = (l^+ + l^-)/2$.

While the obtained voltage dependences of visibility does show
an oscillatory structure similar to experimentally observed lobes,
there are strong differences. Most importantly, in the limit of equal
arms, which is predominantly the experimental situation, 
the visibility calculated in the framework of short-range interaction
model oscillates as 
${\cal V}_G \propto \cos (eV l/v)$ without any decay, since the
non-equilibrium dephasing rate vanishes according to 
the relation~(\ref{eq:At_12}) whatever strong the interaction is. The
situation gets even worse 
when we try to apply the same model to describe experimental data on MZI at
$\nu=1$. Specifically, the model predicts
then a {\it constant} visibility, which is in stark contrast to the
experiments that show lobe structures for $\nu=1$ as well.

To summarize, the Keldysh action approach allows us to determine the
tunneling current through the MZI. The theory includes all the effects
of the interaction, including charging, renormalization, and
dephasing.  The dephasing rate grows linearly with voltage; the 
source of dephasing is the shot noise produced at the
first QPC. 
The calculation of this section were carried out under the assumption
of short-range 
interaction (which would be the case for a system with a nearby
metallic gate). Essential discrepancies between the theoretical findings for
this model and the experimental data indicate the importance of
long-range ($1/r$) character of the Coulomb interaction. The
corresponding model will be studied in the next section.

\section{Long-range interaction}
\label{s4}

In this section, we consider the model of the MZI with the long-range
Coulomb interaction, 
\begin{equation}
 U_0(x)=\frac{e^2}{\epsilon}\frac{1}{\sqrt{x^2+b^2}},
\end{equation}
which is regularized at short distances by the finite width of the
edge state $b$. 
Using the self-consistent electrostatic picture~\cite{Chklovskii92} of
the quantum Hall edge channels in the smooth confining potential, we associate
the scale $b$  with the typical width of a compressible strip. 
The latter satisfies
\begin{equation}
b\gg a\sim l_B,
\end{equation}
where $a$ is the short distance cut-off introduced earlier (see also
estimates of typical experimental parameters in Sec.~\ref{s4.3}). 
The strength of Coulomb interaction is quantified   
by the dimensionless coupling constant 
\begin{equation}
\alpha=\frac{\nu\, e^2}{\epsilon\, \pi v},
\label{eq:alpha}
\end{equation}
where $\epsilon$ is the dielectric constant. As discussed below in
Sec.~\ref{s4.3},  the edge velocity $v$ 
(which is also the velocity of the neutral mode at filling factor
$\nu=2$) is fixed within the self-consistent electrostatic picture  
in such a way that $\alpha\sim 1$.  In the momentum space
\begin{equation}
 U_0(q) = \frac{2 e^2}{\epsilon} K_0(|q|b) \label{eq:U0_Coulomb},
\end{equation}
with the small-$q$ asymptotic behavior
\begin{equation}
U_0(q) \simeq - \frac{2 e^2}{\epsilon}\ln(|q|b d), \quad q\ll b^{-1}, 
\end{equation}
where $K_0$ is the modified Bessel function of the second kind and 
the numerical constant $d =e^{\gamma_E}/2 = 0.89$. 

\subsection{Plasmon correlation function}
\label{s4.1}

Consider now the plasmon phase correlation function~(\ref{eq:J_pf}).
It is given by
\begin{equation}
 \label{eq:PlasmonIntegral}
 J^{>}_{P}(x,t)=\int_0^{\infty} \frac{dq}{q} \Bigr(1-e^{i\phi(q)}\Bigl)e^{-a q},
\end{equation}
where we have introduced the phase $\phi(q)$, containing the $x$- and
$t$-dependence,  
\begin{eqnarray}
\phi(q)&=& q x- \omega_P(q)\,t, \\
\omega_P(q) &=& 
v q\,(1+{\alpha}\,K_0(|q|b)).
\label{eq:w_pq}
\end{eqnarray}
As follows from the integral
representation~(\ref{eq:PlasmonIntegral}), at zero temperature 
$J^{>}_{P}(x,t)$ is an analytic function of time in the upper half-plane. 
We now analyze this function in the relevant 
parameter limit
\begin{equation}
\max\{t, x/v\} \gg b/v \gg a/v.
\end{equation}

\begin{figure}[t]
\includegraphics[width=2.5in]{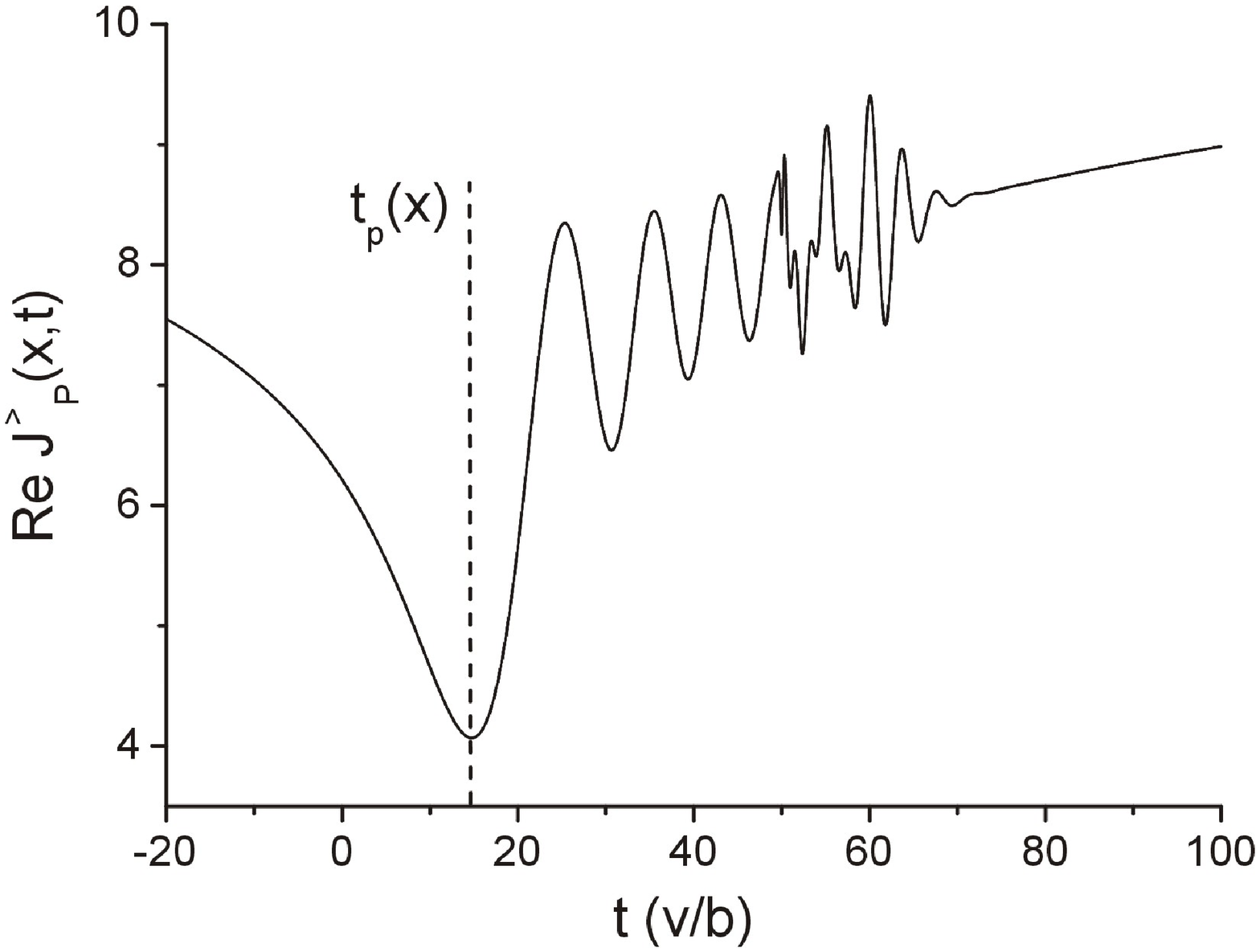}
\includegraphics[width=2.7in]{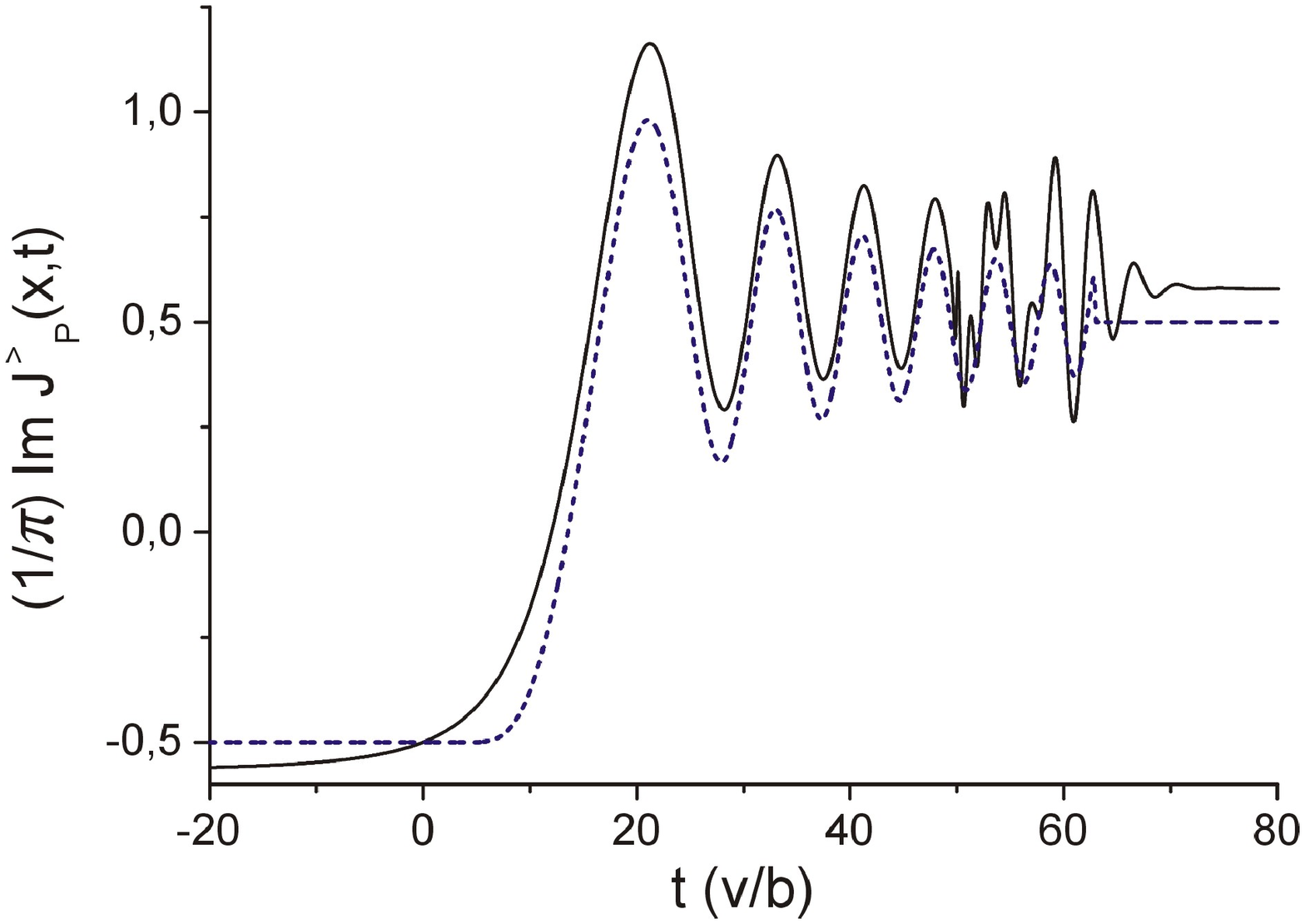}
\caption{Real (upper panel) and imaginary(lower panel)  parts  of the
  plasmon correlation   
function plotted versus the dimensionless time at fixed $x/b=50.0$;
the coupling constant 
$\alpha=1.0$ and $a/b=0.1$. The dashed line shows the analytical
approximation~(\ref{eq:J_pl_asympt}). } 
\label{fig:Jp}
\end{figure}

The time dependence of real and imaginary parts of the phase
correlation function $J^{>}_{P}(x,t)$
for typical values of parameters is shown in Fig.~\ref{fig:Jp}. 
One observes characteristic oscillations which stem from the non-linear
plasmon dispersion $\omega_P(q)=u_p(q)q$ given by Eq.~(\ref{eq:w_pq}).  
As a result, 
the initially localized density wave packet injected into the
interferometer arm  
at some position $x'=0$ and time $t'=0$ is spread after some time $t$,  
when it arrives at given point $x$. This is because its
constituents~--- bosonic modes,   
with different wave vectors $q$ --- propagate with different group velocities.
The real part of the phase correlation function has a pronounced minimum at 
\begin{equation}
t_p(x)= \frac{x}{u_p(q)} \Biggl|_{q\sim 1/x} \simeq 
\frac{x}{\alpha\, v \ln\left( e\,x/b\tilde d \right)},
\label{eq:scale_tp}
\end{equation}
where we have introduced
\begin{equation}
\tilde d(\alpha) = d e^{(\alpha-1)/\alpha},
\end{equation}
and the numerical constant $d$ has been defined below Eq.~(\ref{eq:U0_Coulomb}).
The time $t_p(x)$ is a typical time scale required for the fastest
plasmon (that has the smallest possible momentum $q\sim 1/x$) to 
traverse the distance $x$.
The plots at Fig.~\ref{fig:Jp} should be contrasted with the behavior
of $J^>_P(x,t)$ in the short-range interaction model given by
Eq.~(\ref{eq:J_g_sr}). In the latter case 
the real part of $J^>_P(x,t)$ is a smooth function 
with logarithmic singularity at $t=x/v$,
while the  imaginary part is piecewise constant.

\begin{figure}[t]
\includegraphics[width=2.3in]{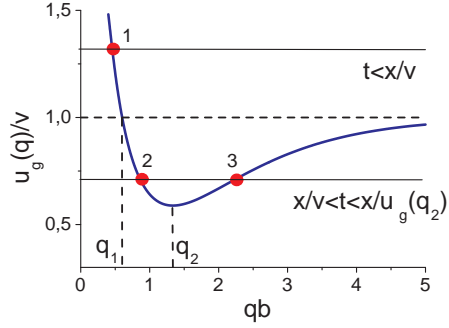}
\caption{The group velocity of plasmons shown as function of the
  dimensionless momentum  
for the coupling constant $\alpha=2$. At $q_1 b = 0.59$ one has $u_g(q_1)=v$
and for $q_2 b = 1.33$ the group velocity is at minimum. }
\label{fig:u_gr}
\end{figure}

The oscillatory behavior of $J_P^>(x,t)$ can be qualitatively
understood using the  
stationary phase approximation to the
integral~(\ref{eq:PlasmonIntegral}). For 
technical details of this analysis we refer the reader to Appendix
D. The saddle point $q_*(x,t)$  
of the phase $\phi(q)$ for given $x$ and $t$ can be found
numerically as the solution of  the equation
\begin{equation}
 x = u_g(q_*) t,
 \label{eq:st_phase}
\end{equation}
where we introduced the group velocity 
\begin{eqnarray}
 u_g(q) &=& \frac{\partial \omega_P(q)}{\partial q} \\ 
 &=&  v \Bigl( 1 + \alpha\, K_0(q b) - \alpha\, (qb) K_1(q b) \Bigr) \nonumber
\end{eqnarray} 
of the plasmon mode. The plot of the function $u_g(q)$
is presented in Fig.~\ref{fig:u_gr}.
As one can see, there are two special momenta, $q_1$ and $q_2$, both
being independent on 
coupling strength $\alpha$. At $q=q_1$ the group velocity matches the
drift velocity $u_g(q)=v$, 
while at $q=q_2$ the group velocity reaches its minimum.
Thus, for  given $x$ and at $t<x/v$ the stationary phase
equation~(\ref{eq:st_phase}) 
has a single root (see Fig.~\ref{fig:u_gr}). In the short-time limit
one finds asymptotically 
\begin{eqnarray}
 q_*(x,t) &\simeq& \left( b \tilde{d} \right)^{-1} e^{-x/\alpha\, v
   t},  \label{eq:phi_star} \\ 
 \phi(q_*) &\simeq& - \alpha \,q_*  v \,t, \qquad t\ll x/v. \nonumber
\end{eqnarray}
The unique saddle point leads to approximately single-period
oscillations in the phase correlation function at 
\begin{equation}
t_p(x) \lesssim t <x/v,
\label{eq:short_t}
\end{equation}
as is indeed seen in the numerical plots in Fig.~\ref{fig:Jp}.
We also remark that at $t\sim t_p(x)$ the optimal phase $\phi(q_*)$
becomes of order unity, thus 
the stationary phase method loses its applicability. At still smaller times
the integral~(\ref{eq:PlasmonIntegral}) is governed by the
contribution of the end point at $q=0$ 
and shows no oscillations. On the other hand, in the time interval
\begin{equation}
x/v < t < x/u_g(q_2)
\label{eq:long_t}
\end{equation}
one finds (see Fig.~\ref{fig:u_gr}) two distinct roots of the stationary 
phase equation~(\ref{eq:st_phase}). In this range the plasmon phase
correlation functions  
acquires characteristic beatings arising from the interference of
the two contributions (with generally incommensurable phases)  
corresponding to two stationary-point momenta.

It is shown in Appendix D that at $t<u_g(q_2)$
the imaginary part of the phase correlation function can
be approximated in the following way:
\begin{equation}
{\rm Im}\,J_P^>(x,t)=\pi/2 - \sqrt{2}\pi{\rm Re}\left[\,{\rm
    Erfc}\sqrt{-i(\phi_*-\phi_0)}\,\right], 
\label{eq:J_pl_asympt}
\end{equation}
where $\phi_*=\phi(q_*(x,t))$ is the stationary phase, 
${\rm Erfc}(z)$ is the complementary error function, 
and the constant $\phi_0$ is defined as the root of the equation
${\rm Re}\left[{\rm Erfc}\sqrt{i\phi_0}\right]= 1/\sqrt{2}$.
The above approximation is shown by the dotted line in
Fig.~\ref{fig:Jp}; one finds a good agreement with 
the exact numerical curve.
We will use this approximation later to evaluate the dephasing action
with the logarithmic accuracy.

\subsection{Visibility}
\label{s4.2}

Let us turn now to the analysis of the visibility. For that purpose we
have to find the differential 
conductance in accordance with general relations~(\ref{eq:GV}) and
(\ref{eq:I}) of Sec.~\ref{s2}. 
We will first perform the analysis in the framework of
the Gaussian approximation
(Sec.~\ref{s2.3}), i.e. neglecting the tunneling 
action ${\rm Im}\, A^*_t$ in Eq.~(\ref{eq:I}). Later we will include the effects
of the non-Gaussian fluctuations.

\subsubsection{ Filling factor $\nu=1$}
\label{s4.2.1}

In the case $\nu=1$ the Green function $G_0$ at zero bias is determined
solely by the plasmon 
part of the total correlation function $J(\xi)$, since the free part
$J_F(\xi)$ cancels against  
the bare Green function $g_0$:
\begin{equation}
 G_0^{>}(x,t)=\frac{1}{2\pi i\, a}e^{-J^>_{P}(x,t)}.
\end{equation}
The ``lesser'' Green function satisfies $G_0^<(x,t) =
\left(G_0^>(x,t)\right)^*$.  
In the case of finite voltage the above expression for the upper arm
is modified to 
\begin{equation}
G_+^\gtrless(x,t)=\exp\left\{-ieV\left(t-\frac{x}{u_{P}(1/L)}\right)
\right\} G_0^\gtrless(x,t), 
\label{eq:G_ne_n1}
\end{equation}
where we took into account the non-equilibrium phase shift~(\ref{eq:MTheta}).

In the following we concentrate on the case of symmetric
interferometer, $l^+ = l^- \equiv  l$. 
We will show that, in contrast to the results of Sec.~\ref{s3},
even in this situation the visibility is a non-trivial function of voltage. 
A small mismatch in the lengths of upper and lower arms will not affect our 
conclusions essentially.  

\begin{figure}[t]
\includegraphics[width=2.8in]{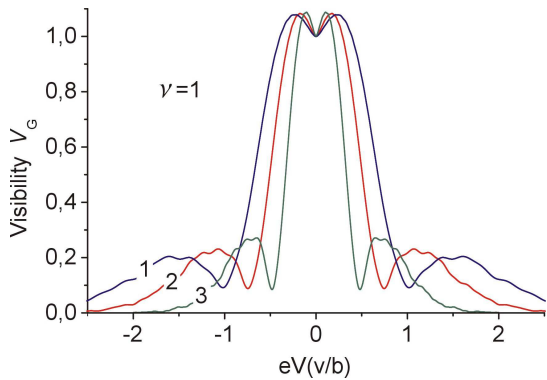}
\caption{Visibility of Aharonov-Bohm oscillations in a symmetric
  Mach-Zehnder interferometer at $\nu=1$  
calculated in the model of long-range Coulomb interaction with the
coupling constant $\alpha=1$, 
shown for $a/b=0.1$ and different ratios of $l/b$: (1)-30; (2)-50; (3)-100.}
\label{fig:Vis_n1_L}
\end{figure}

In the Gaussian approximation the visibility of interference oscillations
can be found by integrating  Eq.~(\ref{eq:I})  
over the time numerically. As discussed above, we neglect the dephasing caused
by the shot noise at this stage and will restore it later.  
Using Eqs.~(\ref{eq:Dt}), we obtain for the delay time
\begin{equation}
 \Delta t \,=\, \frac{l}{u_{P}(1/L)}\,\simeq\, t_p(l)\,.
 \label{eq:Dt_n1}
\end{equation}
In a typical experimental layout the distance from source to
drain $L$ is of the order of MZI size  $l$. Thus, 
within the logarithmic accuracy (i.e. up to a numerical factor in the
argument of the logarithm) the delay time is actually equal to the 
scale $t_p(l)$ defined by Eq.~(\ref{eq:scale_tp}).

\begin{figure}[b]
\includegraphics[width=2.6in]{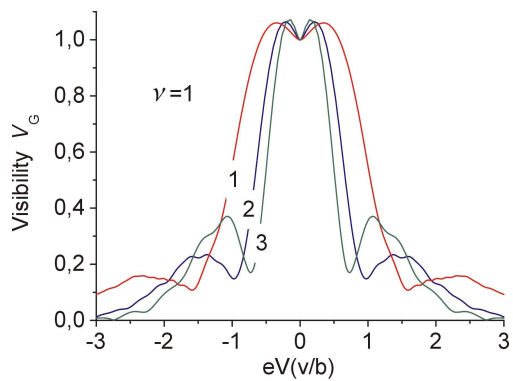}
\caption{Visibility of Aharonov-Bohm oscillations in a symmetric MZI at $\nu=1$ 
calculated in the model of long-range Coulomb interaction for
different coupling constant $\alpha$: 
(1)-2.0; (2)-1.0; (3)-0.5. The other parameters are:  $l/b=30$ and $a/b=0.1$.}
\label{fig:Vis_n1_a}
\end{figure} 

The visibility ${\cal V}_G$ calculated in accordance 
with the definition~(\ref{eq:Vis_def}) is shown in Fig.~\ref{fig:Vis_n1_L} for
different values of the ratio $l/b$
and in Fig.~\ref{fig:Vis_n1_a} for different values of 
the coupling constant $\alpha$.
In all cases the MZI is symmetric, meaning that $l^+=l^-$ and both
QPCs are equivalent. 
The plots clearly show a non-trivial voltage dependence of visibility,
with three lobes. The width $\epsilon_0$ of the central peak  
scales as 
\begin{equation}
 \epsilon_0 \sim 2\pi/t_p(l) \sim 2\pi\alpha \frac{v}{l} \ln(l/b)
 \label{eq:Epsilon_0}
\end{equation}  
A dip in the visibility at small voltage is the result of a zero-bias
anomaly in the  
direct ${\cal G}_0(V)$ and interference ${\cal G}_\Phi(V)$ part of conductance.

As has been already emphasized, the appearance of the lobe structure
in visibility   
in the model of long-range Coulomb interaction at $\nu=1$ crucially
depends on non-linearity of plasmon dispersion, which gives rise to
the oscillatory behavior of the phase correlator $J_P(x,t)$ as the function of
time. We also note that  
additional weak oscillations in ${\cal V}_G$ (seen on top of the
lobe pattern) 
stem from the beating in the plasmon correlation function in the
time interval~(\ref{eq:long_t}), which is due to plasmons with short
momenta $q\sim b^{-1}$.   

Our calculations in this subsection are closely related to those of
Ref.~\onlinecite{Chalker07} where the effect of non-linearity of plasmon dispersion
on properties of a $\nu=1$ MZI was analyzed. The crucial difference is
that we also take into account the phase shift~(\ref{eq:MTheta}) representing
a non-equilibrium charging effect. For this reason, our
non-equilibrium Green function~(\ref{eq:G_ne_n1}) and the 
resulting visibility ${\cal V}_G$ differ from their counterparts in
the work~\onlinecite{Chalker07}. The voltage dependence of visibility shown
in Figs.~\ref{fig:Vis_n1_L} and \ref{fig:Vis_n1_a} results from a
combined effect of non-linear dispersion and charging.

\subsubsection{Filling factor $\nu=2$}
\label{s4.2.2}

\begin{figure}[t]
\includegraphics[width=2.8in]{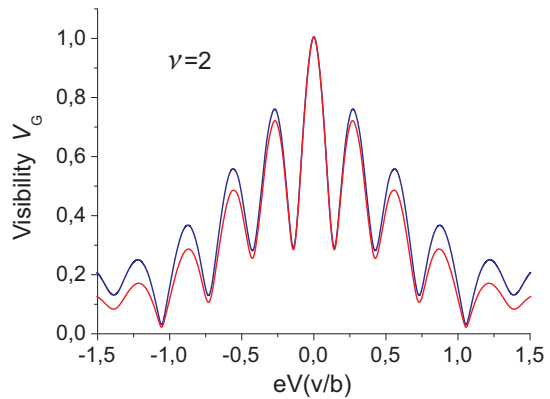}
\caption{Visibility of Aharonov-Bohm oscillations in a symmetric MZI at $\nu=2$ 
calculated in the model of long-range Coulomb interaction with the
coupling constant $\alpha=1$, 
shown for $l/b=30$ and $a/b=0.1$. Upper curve: Gaussian approximation; 
lower curve: shot-noise dephasing with $\tau_1^2=0.2$ is taken into account.}
\label{fig:Vis_n2_L30}
\end{figure}

We have carried out analogous calculations for the setup with $\nu=2$.
As before, we concentrate on the symmetric MZI.  
In the case of two edge states per interferometer arm a tunneled
electron excites plasmon and neutral modes.  
The electron propagator at zero voltage acquires the form 
\begin{equation}
 G_0^{>}(x,t)=\frac{1}{2\pi\sqrt{ i a v}}\frac{e^{-J^>_{P}(x,t)/2}}{\sqrt{x/v- t + i a/v}}\,. 
\end{equation}
The delay time for the setup of Fig.~\ref{MZI_scheme} with one of
outer channels biased can be found using Eq.~(\ref{eq:Dt}),
\begin{equation}
 \Delta t \simeq \frac{l}{2}\left(\frac{1}{u_{P}(1/l)}+\frac{1}{v}\right)\,.
\end{equation}
It is instructive to compare this result with Eq.~(\ref{eq:Dt_n1}):
one sees a clear manifestation of two modes (plasmon and neutral) for
$\nu=2$ instead of a single plasmon mode at $\nu=1$. 
The voltage dependence of visibility calculated
numerically in the Gaussian approximation is shown 
by the upper curve in Fig.~{\ref{fig:Vis_n2_L30}}. 
(The lower curve in the same plot is the visibility calculated 
in the instanton approximation, which
takes into account the extra dephasing due to the non-equilibrium
shot-noise, Sec.~\ref{s4.2.3}.)
One can see that the voltage dependence of visibility 
is qualitatively different from the $\nu=1$ setup.
Specifically, 
the visibility ${\cal V}_G(V)$ shows many oscillations (``lobes'') as the
function of bias with a typical width given by the Thouless energy of the MZI,
\begin{equation}
\epsilon_{\rm Th}=\pi v/l \,.
\label{eq:Th_scale}
\end{equation}
This energy scale differs from the characteristic scale $\epsilon_0$
characterizing lobes in the $\nu=1$ MZI
[see Eq.~(\ref{eq:Epsilon_0})] by a logarithmic factor. 

The results of this subsection should
be also contrasted with those of Sec.~\ref{s3}. 
The visibility ${\cal V}_G(V)$ calculated there
for the same setup ($\nu=2$) but in the model of short-range interaction
demonstrates the $\cos$-like oscillations which, however, 
do not decay in the case of a symmetric MZI. 
In the $1/r$ model of Coulomb interaction considered here, 
the visibility ${\cal V}_G(V)$ oscillations do decay already in the
Gaussian approximation, as seen in Fig.~\ref{fig:Vis_n2_L30}. 
Furthermore, as we are going to show in Sec.~\ref{s4.2.3}, the
shot-noise dephasing leads in this case to an additional suppression of
oscillations of visibility and to a decay of visibility down to zero 
at high voltage (rather than to a constant as in the Gaussian model).

\subsubsection{Non-equilibrium dephasing by the shot noise}
\label{s4.2.3}

Let us now calculate the suppression of the visibility due to
non-Gaussian phase fluctuations  
following the method of Sec.~\ref{s2.4}. 

In the instanton approximation the optimal phases 
$\Theta^{*,a}_k$ are expressed in terms of the phase correlation function
$J(x,t)$ according to Eq.~(\ref{eq:Phi_Inst}). 
We saw in section \ref{s3} that the special
properties~(\ref{eq:SymmetryPropertiesJ}) of the causal  
correlators $J^{T/\tilde T}(x,t)$ in the case of short-range
interaction greatly simplify the calculations.  
The question arises whether these properties are also valid in the
case of long-range interaction. Since
$J^{>}(x,t)=\left(J^{<}(x,t)\right)^*$, we conclude that $J^{>}$ and 
$J^{<}$ may differ only in the imaginary part.
As is clearly seen in Fig.~\ref{fig:Jp},
at $x>0$ and $t<0$ (or, at $x<0$ and $t>0$) we have ${\rm Im}
J^{>}_{P} \simeq {\rm Im} J^{>}_{F}=-\pi/2$, so that the imaginary
part of the full correlator, ${\rm Im} J^{>}$, in the same range of
parameters is almost zero. It follows that the relations
(\ref{eq:SymmetryPropertiesJ}) are approximately satisfied. This is
a manifestation of the fact that even in the presence of spectral dispersion
the edge modes retain their chiral character.  

Performing calculation similar to those in Sec.~\ref{s3}, 
we arrive again at the result (\ref{eq:Im_At_12}), 
which can be cast in the form
\begin{equation}
 \mathrm{Im}\,\mathcal{A}_{t,12}^*= 2\tau_1^2 \mathcal{P}(eV) \Upsilon (l),
 \label{eq:ImA_t12_LR}
\end{equation}
where $\mathcal{P}(eV)$ is the noise power,
\begin{equation}
\mathcal{P}(eV)=-\int_{-\infty}^{\infty}
dt\left(\tilde{\Pi}_{11}^{<}(t)+\tilde{\Pi}_{11}^{>}(t)\right),
\end{equation}
and 
\begin{equation}
\Upsilon (l) = \int_0^{+\infty} dt'
\sin^2\left(\frac{2}{\nu}\mathrm{Im}\, J^>_{\rm P}(l,t') -
  \frac{\pi}{\nu}\right)  
\label{eq:Bdwell_time}
\end{equation}
is a characteristic bosonic dwell time related to the size of the MZI.
In what follows, we will find the dephasing action~(\ref{eq:ImA_t12_LR})
with the logarithmic accuracy (up to a coefficient of order unity
under the logarithm).

First, we evaluate the noise, using its representation in terms of
interacting Green function 
\begin{equation}
\label{eq:Pol}
\mathcal{P}(eV)=4v^2\int_{-\infty}^{+\infty} dt \sin^2(eVt/2)\, {\rm Re}
\left[G_0^{>}(0,t)\right]^2. 
\end{equation}
To derive this expression, we have used the relation
$G_0^<(0,t)=\left(G_0^>(0,t)\right)^*$ 
and the fact that at zero temperature and at $eV\to 0$ the
shot noise should vanish.
To find the interacting Green function, we evaluate the momentum 
integral~(\ref{eq:PlasmonIntegral}). At large times $t\gg b/v$ it
is given by 
\begin{equation}
 J^>_P(t)\simeq \ln(1/q^*)  + {\rm const} \,.
\end{equation}
Here the momentum $q^*$ satisfies the condition $|\phi(q^*)|\sim 1$,
which at $x=0$ leads to 
\begin{equation}
 (q^*)^{-1} \simeq \alpha vt \ln\left( {e v t}/{b\, \tilde d}\right)
\end{equation}
(here $e$ stands for the base of natural logarithms rather than for the electron charge).
On the other hand, at small times $t\ll b/v$ the plasmon correlator is
determined by the 
large-momentum behavior of the integrand in Eq. (\ref{eq:PlasmonIntegral}),
where interaction is absent and thus $J^>_P(t)$ should coincide with the
free result.  An approximation for the interacting Green function
which respects both limits is given by
\begin{equation}
G^>_0(0,t)\sim\frac{1}{2\pi \alpha^{1/\nu}(v t - ia)
  \ln^{1/\nu}\left( e + i e v t/b\,\tilde d\right)},  
\end{equation}
where we have also taken into account that $G^>_0(0,t)$ must be
analytic in the upper half plane of complex $t$. 
The main contribution to the integral in Eq.~(\ref{eq:Pol}) comes from
times $t\sim 1/eV$.  
Evaluating the integral, we finally find
\begin{equation}
 \mathcal{P}(eV)\sim\frac{|eV|}{2\pi\alpha^{2/\nu}\ln^{2/\nu}\Bigl(e +
   \,v/|eV|b\,\tilde d\Bigr)}.
 \label{eq:NoiseV}
\end{equation}
Note that the noise power reveals a weak zero-bias anomaly in the
chiral 1D system, which 
originates from the long-range nature of the Coulomb interaction.

Let us now evaluate the bosonic dwell time~(\ref{eq:Bdwell_time}). One
can employ the  
saddle point approximation~(\ref{eq:J_pl_asympt}) and change the
integration variable from time 
$t'$ to phase $\phi$ using the relation~(\ref{eq:phi_star}). We find
with the logarithmic accuracy 
\begin{equation}
 t \simeq \frac{l}{\alpha\, v}\left[\ln\left(\frac{l}{b\,\phi\, \tilde
       d} \right)\right]^{-1} \,. 
\end{equation}
This enables us to cast the dwell time~(\ref{eq:Bdwell_time}) into the form
\begin{eqnarray}
\Upsilon (l)&\simeq&\frac{l}{\alpha v}\int_0^{\phi_{\max}} 
\frac{d\phi}{\phi}\ln^{-2}\left(\frac{l}{b\,\phi\, \tilde d} \right)
\nonumber \\ 
&\times&\sin^2\left[\frac{2\pi\sqrt{2}}{\nu}\mathrm{Re}
\left( {\rm Erfc}\sqrt{i(\phi_0+\phi)}\right)
\right],
\label{eq:Dl}
\end{eqnarray}
where $\phi_{\max} \sim \alpha (l/b)/e$. This upper limit of
integration originates from the fact that the integral 
in Eq.~(\ref{eq:Bdwell_time}) is actually limited to times $t \lesssim
l/v$. Let us now assume that the interaction is not too weak, so
that $\phi_{\max} \gg 1$. This is always the case 
in the experiment, as will be shown in Sec.~\ref{s4.3}. Since the
integral~(\ref{eq:Dl}) converges at large $\phi$ (and is dominated by
$\phi\sim 1$), 
one can now extend the integration up to infinity and
also neglect the weak logarithmic dependence 
on the phase $\phi$. In this way we find 
\begin{equation}
 \Upsilon (l) \simeq C_\nu \left(\frac{l}{\alpha v}\right){\ln^{-2}
   \left(\frac{l}{b\, \tilde d} \right)},  
 \label{eq:Upsilon_l}
\end{equation} 
where $C_1=3.96$ and $C_2=2.97$. 

Equation (\ref{eq:ImA_t12_LR}) together with Eqs.~(\ref{eq:NoiseV}) and
(\ref{eq:Upsilon_l})
is the main 
result of this subsection. We see that the non-equilibrium dephasing
action in the model of  
long-range Coulomb interaction is a sub-linear function of both the
voltage and the size of  
the interferometer.  Equation (\ref{eq:ImA_t12_LR}) tells us that the
non-equilibrium dephasing in the MZI occurs  
due to the intrinsic shot noise. This noise is transferred by plasmons
(in case of $\nu=1$) or by a 
superposition of plasmon and neutral bosonic modes (in case of
$\nu=2$), which typically propagate 
through the MZI faster than electrons, as is seen from
Eq.~(\ref{eq:Upsilon_l}). 

In Fig.~\ref{fig:Vis_n2_L30} we illustrate the suppression of the
visibility by the shot noise for the setup with $\nu=2$. 
The effect of shot-noise dephasing leads to suppression of visibility
at high voltages. However, the lobe structure remains well preserved:
the dephasing effect becomes strong only for voltages much larger than
the period of the oscillations in visibility. The reason for this is
twofold. First, this is a factor $\tau_1^2$ in
Eq.~(\ref{eq:ImA_t12_LR}) that was assumed to be small in our
calculation. (We used the value $\tau_1^2=0.2$ in the plot.) 
We can expect a somewhat stronger  
dephasing due to the shot noise $\sim \tau_1^2(1-\tau_1^2)$,  
when the QPC transparency is of order $50\%$.
Second, there are logarithmic factors in Eqs.~(\ref{eq:NoiseV})
and (\ref{eq:Upsilon_l});  their combined effect at the scale $eV \sim v/b$ and at $\nu=2$ leads to 
the suppression of the dephasing action by a factor $1/\ln^3(l/b) \ll 1$. Because of this
factor, the lobe structure should not be destroyed by dephasing
(i.e. many oscillations will be seen) even in the case of most noisy QPC with
transparency of order $50\%$, in agreement with experiments.

A more detailed comparison of our findings with experiment is
presented below. 

\subsection{Comparison to experiment}
\label{s4.3}

Let us now discuss the relation of theoretical results of Sec.~\ref{s4}
to experimental observations. 
The striking lobe-type structure in the visibility as function of
voltage at low temperature 
was discovered for the first time in the experiment~\cite{heiblum06}.
At filling factor $\nu=1$ three lobes were reported, while
at $\nu=2$ five lobes were observed. Later experiments of the same
group with the use of somewhat different layout \cite{Ofek}
showed up to 9 lobes for $\nu=2$.  The important
part of the MZI setup in the 
experiment~\cite{heiblum06} was an additional QPC0, which enabled to
apply different chemical  
potentials to the outer and inner edge channels, see Fig.~\ref{MZI_scheme}. 
These observations were corroborated by subsequent
works~\cite{roche07,strunk08}, where an  
analogous MZI layout with QPC0 was studied. 

Our theoretical calculations
carried out within the model of long-range Coulomb interaction (Sec.~\ref{s4})
compare well with the above experimental findings.
On the qualitative level, we find three lobes in the case of $\nu=1$ 
(Figs.~\ref{fig:Vis_n1_L} and \ref{fig:Vis_n1_a}) and structure with
many lobes for $\nu=2$ (Fig.~\ref{fig:Vis_n2_L30}), in agreement with
the experiments.

Let us also note that a more complicated behavior of the finite-bias
visibility was observed in another 
experimental work~\cite{schoenenberger}, where the MZI setup did not
contain an additional QPC0. 
The emphasis of this study was on the asymmetry of the MZI
characteristics when the transparency of the QPC1 
was varied from 0 to 1. We cannot treat this feature within our
approximation valid in the weak-tunneling regime. 

More quantitative comparison of theoretical and experimental results
requires the knowledge 
of the edge velocity $v$ and the range $b$ of the Coulomb
interaction. To estimate them, we
use the results of works~\onlinecite{Chklovskii92, Aleiner94}.
In Ref.~\onlinecite{Aleiner94} the excitation spectrum of the compressible
Hall liquid in the classical 
limit $\nu \gg 1$ was found. It consists of the magnetoplasmon mode with
\begin{equation}
\omega_0(q) = v q \ln\left( e^{-\gamma}/|q| \bar l \right),
\end{equation}
and the acoustic spectrum with 
\begin{equation}
\omega_n(q) = v q/n, \quad n\ge 1,
\label{eq:omega_0}
\end{equation} 
where $v$ is the drift velocity, 
\begin{equation}
 v = \frac{\nu e^2}{\epsilon\pi\hbar},
 \label{eq:v_drift}
\end{equation} 
and $\bar l$ is the half of the width of the depletion layer between
the confining gate and the 2DEG. The latter scale was found in
Ref.~\onlinecite{Chklovskii92} and is controlled by the gate voltage,
\begin{equation} 
\bar l \simeq \frac{V_g \epsilon}{4\pi n_0 e}.
\end{equation}
For the typical concentration $n_0 \sim 2\times 10^{11}$cm$^{-2}$, the
gate potential $V_g \sim $1V,
and $\epsilon=12.5$, one thus gets an estimate $\bar l \sim 110\,$nm. 
This result is also believed to be applicable 
in the case of etched mesostructures, e.g. those used in the MZI
setups. In this case the voltage $V_g$ 
should be associated with a work function, which for GaAs-AlGaAs
heterostructures has the same energy scale $\sim 1$~eV.

In the quantum Hall regime, at $\nu=1$ and $\nu=2$, 
one expects that the above modes with $n=0,1$ should match the two
chiral bosonic modes of our theory. 
We thus see that the drift velocity~(\ref{eq:v_drift}) would fix the
interaction constant $\alpha$, 
defined by Eq.~(\ref{eq:alpha}), to
be of order 1. At the same time, comparing the dispersion
relation~(\ref{eq:omega_0})  
to Eqs.~(\ref{eq:U0_Coulomb}) and (\ref{eq:w_pq}), we obtain the
estimate for the scale $b$ in our theory, $b \simeq 2e^\gamma \bar l
\sim 400\,$nm.   

Let us first discuss the case $\nu=2$. 
For the typical size of the MZI, $l \sim 10\,\mu$m, one gets the ratio
$l/b \sim 25$. 
This parameter controls the overall number of lobes in our theory, 
$N\sim l/\pi b$, cf. Fig.\ref{fig:Vis_n2_L30}. 
For sufficiently large transparency of QPCs the shot-noise dephasing
will suppress the visibility at sufficiently large voltages
corresponding to lobe indices $\gtrsim 1/\ln^3(l/b)$, see 
Sec.~\ref{s4.2.3}. However, for $l/b \sim 25$ this happens to be not
so important for the lobe structure. To find the energy scale for the
lobe structure, we estimate the drift velocity according to
Eq.~(\ref{eq:v_drift}) that yields $v\sim 10^5$~m/s.  
This results in the following value of the Thouless energy
~(\ref{eq:Th_scale}):  $\epsilon_{\rm Th}\sim 20\,\mu$eV. 
This is indeed the typical energy scale seen 
in the above discussed experiments~\cite{heiblum06,roche07,strunk08}.

We turn to the case of filling factor $\nu=1$. We obtain bias
dependences of visibility with three lobes in a sufficiently broad
range of parameters (see
Figs.~\ref{fig:Vis_n1_L} and~\ref{fig:Vis_n1_a}), in agreement with
experimental works~\cite{heiblum06,strunk08}. The energy scale for the
lobes in our theory is given by Eq.~(\ref{eq:Epsilon_0}), which is larger 
than the Thouless energy $\epsilon_{\rm Th}$ by a factor $\ln
(l/b)$. For realistic values of parameters (see above), this factor is
$\sim 3$. In the experiment the energy scales for $\nu=1$ and $\nu=2$
are practically equal. This discrepancy may be partly explained by the
fact that, according to Eq.~(\ref{eq:v_drift}), the drift velocity is
expected to be twice smaller for $\nu=1$ compared to $\nu=2$, which
reduces the energy scale by factor of 2, partly compensating the
logarithmic factor.

\section{Summary}
\label{s5}

In this paper we have discussed the influence of the Coulomb
interaction on the quantum coherence  
in the electronic Mach-Zehnder interferometer (MZI) formed by integer
quantum Hall edge states at filling 
fractions $\nu=1$ and $\nu=2$ out of equilibrium. Our main results can
be summarized as follows.

\begin{enumerate}

\item
We have developed the non-equilibrium functional bosonization 
framework which enables us to build up the Keldysh action of
interacting electrons in the MZI. The most non-trivial term in the 
action is expressed 
in terms of a single-particle time-dependent interferometer scattering
matrix in the dynamically fluctuating field and has a structure of a
Fredholm determinant similar to those appearing in the theory of full
counting statistics.  
We have used this action to analyze the limit of weak electron
tunneling between interferometer arms in the case  
of arbitrarily strong e-e interaction. Our theory
contains all interaction-induced effects on transport through MZI,
including charging, dispersion non-linearity, and non-equilibrium
decoherence.   

\item
Restricting at first the theoretical analysis to the Gaussian approximation 
for electron phase fluctuations, we have readily reproduced the previous theoretical
results related to oscillations  (lobe structure) 
in the dependence of the visibility on the bias
voltage. Going beyond the Gaussian approximation, we have used a real-time instanton approach
to evaluate the non-equilibrium dephasing rates that lead to 
further suppression of the Aharanov-Bohm oscillations in
conductance with the increase of voltage. 
We have found that the out-of-equilibrium dephasing rate is
proportional to the voltage dependent  
shot noise of the first quantum point contact (QPC1),
defining the MZI, and originates from  
the emission of non-equilibrium plasmons and neutral bosonic modes in
course of inelastic electron tunneling.

\item
The results obtained within the model of short-range interaction
(Sec.~\ref{s3}) show strong contradictions to the
experiment. Specifically, in the case of equal arms the visibility
oscillations in the $\nu=2$ interferometer do not decay with
voltage. For $\nu=1$ the problem is even more severe, as the
visibility does not depend on the voltage at all.

\item
Considering the realistic model of the long-range ($1/r$) Coulomb
interaction, we are able to explain the experimentally observed 
dependence of the visibility  
of the interference signal at filling fraction $\nu=1$ and $\nu=2$.  
The origin of this effect is found to be a combination of three factors: 
\begin{itemize}
\item[(i)] the electrostatic phase shift effect, related to 
the charge imbalance on different arms of the interferometer,
\item[(ii)] the interaction induced effect of the
plasmon dispersion, and 
\item[(iii)] the out-of-equilibrium decoherence due to the intrinsic
  shot noise.  
\end{itemize}
Using realistic parameters, we find 
three lobes in the bias dependence of visibility in the case of $\nu=1$
(Figs.~\ref{fig:Vis_n1_L} and \ref{fig:Vis_n1_a})  and structures with
many lobes for $\nu=2$ (Fig.~\ref{fig:Vis_n2_L30}), in agreement with
the experiments. 
The energy scale for the lobe structure in the
visibility for $\nu=2$ is given by the Thouless energy of the MZI
which is estimated as $\sim 20\,\mu$eV for realistic parameters, again
in good agreement with experiment. For $\nu=1$  the energy scale for the
lobes in our theory is enhanced by a factor $\ln (l/b) \simeq 3$,
which is partly compensated by a difference in the drift velocity 
(and thus Thouless energy) at $\nu =1$ and $\nu=2$. 
There remains some discrepancy in this point with the
experiment which indicates that the energy scales for $\nu=2$ and
$\nu=1$ lobe structures are practically equal. This issue may be worth
further study.

\end{enumerate}

We anticipate that the approach developed in this work will be useful
for a much broader class of electronic interference setups relevant to
current or forthcoming experiments. The prospects for future research 
include, in particular, interferometers operating in the fractional
quantum Hall regime as well as those built on edge states of
topological insulators.

\section{Acknowledgements}
\label{s6}

We thank Y.~Gefen, I.V.~Gornyi, M.~Heiblum, I.P.~Levkivskyi, S.~Ngo Dinh, N. Ofek, and D.G.~Polyakov for
useful discussions.  
This work was supported by the German-Israeli Foundation Grant No.\ 965,
by the EUROHORCS/ESF EURYI Awards scheme ( Project ``Quantum
Transport in Nanostructures" ) and by CFN/DFG. 

\appendix

\section{Regularization of the functional determinant}

In this appendix we derive an exact expression for the
tunneling action  
${\cal A}_t = {\cal A} - {\cal A}_0$ 
where ${\cal A}$ is the action given by Eq.~(\ref{eq:Levitov}) and
${\cal A}_0$ is its value in the absence of tunnel coupling between
the interferometer arms. 

 Let us denote the first (determinant) term of the action
 (\ref{eq:Levitov}) as
\begin{equation}
i\mathcal{A}_c = \ln \det \left[1+(S_b^{\dagger}e^{i\hat{\chi}}S_f
  -1)\hat{f} \right]. 
\end{equation}
Then in the absence of tunneling we have 
\begin{equation}
 i\mathcal{A}_c^{(0)}=\ln\det\left[1+(e^{-i\tilde{\psi}_b}e^{i\hat{\chi}}e^{
-i\tilde{\psi}_f}-1)\hat{f}\right] \, ,
\end{equation}
where 
\begin{equation}
\tilde{\psi}^{\pm}_{f/b}=-\frac{1}{v_F} \int_0^{L^{\pm}} dx'
\varphi^{\pm}_{f/b}(x',t+x'/v_F)
\end{equation}
are the phases collected along the way from $0$ to $L^{\pm}$ without tunneling. 
For the full distribution function $\hat{f}$ one can write:
\begin{equation}
\hat{f}(t_1,t_2)=e^{-i\varphi_L(t_1)}f_F(t_1-t_2)e^{i\varphi_L(t_2)}, 
\label{f_gauge_transform}
\end{equation}
where $\varphi_L^{\pm}(t)=eV^{\pm}t$, and $f_F$ is the Fermi distribution
function in time domain at zero voltage. Since the gauge
transformation (\ref{f_gauge_transform}) does not affect the determinant,
we can write $i \mathcal{A}_c = \ln \det M$ and  $i
\mathcal{A}_c^{(0)} = \ln \det  M_0$, with
\begin{eqnarray}
 M&=&f_>+e^{i\varphi_L} S_b^{\dagger} e^{i\hat{\chi}} S_f e^{-i\varphi_L} f_<
\nonumber \\
 M_0&=&f_>+e^{-i\tilde{\psi}_b} e^{i\hat{\chi}}  e^{i\tilde{\psi}_f} f_< \, .
\end{eqnarray}
Here $f_<=f_F$ and $f_>=1-f_F$, as explicitly defined by Eq.~(\ref{eq:f_gl}). 
Note that after the gauge transformation $M_0$ does not depend on voltage.
The tunneling action can be now represented in the form 
\begin{equation}
i \mathcal{A}_t = \ln \det M M_0^{-1}\,,
\end{equation}
and the next step is to invert $M_0$. 

To perform the inversion, we employ the projection properties of
the Fermi distribution function at zero temperature. This approach is
well known from the works on Fermi edge singularity \cite{Nozieres}
and its recent generalization on the matrix case~\cite{dAmbrumenil05}.
In the energy domain, $f_<$ is the projector on occupied states, whereas
$f_>$ projects on 
unoccupied states. Therefore, we have $f_>^2=f_>$,  $f_<^2=f_<$, $f_> f_<=0=f_<
f_>$, and $f_<+f_>=1$. The same relations hold in the time domain as well, where
the product of two operators is understood in the sense of
convolution. Let now $\psi^{\vee}(t)$ 
($\psi^{\wedge}(t)$) be a function, which is analytic in the lower (upper)
complex half-plane. Then the following relation hold:
\begin{eqnarray}
 \label{eq:projection}
  f_< e^{i \psi^{\vee}} f_< &=& f_< e^{i \psi^{\vee}},  \qquad  f_>e^{i
\psi^{\vee}} f_> = e^{i \psi^{\vee}}f_>\nonumber\\
  f_< e^{i \psi^{\wedge}} f_< &=& e^{i \psi^{\wedge}}f_<, \qquad f_>e^{i
\psi^{\wedge}} f_> = f_>e^{i \psi^{\wedge}}\, .
\end{eqnarray}
For instance, let us proof the first relation:
\begin{eqnarray}
 & &\left(f_< e^{i\psi^{\vee}} f_<\right)(t_1,t_2)\nonumber\\
 &=&\frac{1}{(2\pi i)^2}\int dt \frac{1}{t_1-t+i0}
e^{i\psi^{\vee}(t)}\frac{1}{t-t_2+i0}\nonumber\\
 &=&-\frac{1}{2\pi i} \frac{e^{i\psi^{\vee}(t_2)}}{t_1-t_2+i0}=f_<(t_1-t_2)
e^{i\psi^{\vee}(t_2)}\, .
\end{eqnarray}
Due to analytical properties of $\psi^{\vee}(t)$ this integral is
defined by a single  
residue at $t=t_2-i0$.

Furthermore, for any function $A(t)$ with the support at real time $t$, 
\begin{equation}
A^{\wedge}(t)=\int dt' f_<(t-t') A(t')
\end{equation}
is the analytic function of complex $t$ in the upper half-plane, and
\begin{equation}
A^{\vee}(t)=\int dt' f_>(t-t') A(t') 
\end{equation}
is the analytic function in the lower half plane, respectively.

With the help of these zero temperature projection properties, we readily obtain
the inverse of $M_0$. First, let us introduce the quantum component of
the field $\tilde{\psi}$ as 
\begin{equation}
\tilde{\psi}_q=\tilde{\psi}_f-\tilde{\psi}_b. 
\end{equation}
Then we define 
\begin{equation}
\psi^{\wedge}=f_<\tilde{\psi}_q+\hat{\chi}/2,
\end{equation}
which is analytic in the upper
half-plane, and 
\begin{equation}
\psi^{\vee}=-f_>\tilde{\psi}_q-\hat{\chi}/2, 
\end{equation}
which is analytic in the lower half-plane.
Therefore we have 
\begin{equation}
\psi^{\wedge}-\psi^{\vee}=\tilde{\psi}_q+\hat{\chi}, 
\end{equation} and thus can write
\begin{equation}
 M_0=f_>+e^{-i\psi^{\vee}}e^{i\psi^{\wedge}}f_<=e^{-i\psi^{\vee}}\left[e^{i\psi^
{\vee}}f_>+e^{i\psi^{\wedge}}f_<\right] .
\end{equation}
The inverse of this operator is given by
\begin{equation}
 M_0^{-1}=\left[e^{-i\psi^{\vee}}f_>+e^{-i\psi^{\wedge}}f_<\right]
e^{i\psi^{\vee}}\, ,
\end{equation}
which can be easily checked, using the relations
(\ref{eq:projection}). Similarly, 
we calculate $M M_0^{-1}$ and find for the tunneling action:
\begin{equation}
 i\mathcal{A}_t=\ln \det \left[ f_>+e^{i\psi^{\vee}+i\varphi_L} S_b^{\dagger}
e^{i\hat{\chi}} S_f e^{-i\varphi_L-i\psi^{\wedge}} f_<\right] \, .
\label{eq:At_app}
\end{equation}
Let us now introduce the extra gauge phase
\begin{equation}
\hat{\lambda}=f_> \tilde{\psi}_f + f_< \tilde{\psi}_b. 
\label{eq:lambda1}
\end{equation}
It enable us to rewrite $\psi^{\wedge\vee}$ in the form 
\begin{eqnarray}
  \psi^{\wedge} &=& \tilde{\psi}_f-\hat{\lambda}+\hat{\chi}/2, \nonumber \\
  \psi^{\vee}&=&\tilde{\psi}_b-\hat{\lambda}-\hat{\chi}/2. 
\end{eqnarray}
Substituting them into Eq.~(\ref{eq:At_app}) we finally obtain the
tunneling action 
in the form~(\ref{eq:At_main}) as stated in the main body of the paper.

\section{Matrix elements of $Q$}

We define the $Q$-matrix as 
\begin{equation}
Q=e^{i\tilde{\psi}_b} S_b^{\dagger}(\chi) S_f(\chi)e^{-i\tilde{\psi}_f} = 
\left(
\begin{array}{cc}
R & T \\
T' & R'
\end{array}
\right),
\end{equation}
see Eq.~(\ref{eq:Q}) of the main text. 

To evaluate the action ${\cal A}_t$ in the weak tunneling limit up to the terms 
of order of ${\cal O}(\tau^2)$, one needs to know the matrix elements
$Q_{\mu\nu}$ with the following accuracy:  
\begin{eqnarray}
 R&=&1-\tau_1^2-\tau_2^2 - \tau_1 \tau_2 \left\lbrace
\Gamma_1^{\dagger b} \Gamma_2^b e^{i \Phi} + \Gamma_2^{\dagger f} \Gamma_1^f
e^{-i \Phi}\right\rbrace\nonumber \\
    & & +\,\tau_1 \tau_2 \left\lbrace \Gamma_1^{\dagger b} \Gamma_2^f e^{i \Phi} +
\Gamma_2^{\dagger b} \Gamma_1^f e^{-i \Phi}\right\rbrace  e^{-i \chi} \\
    & & +\, \tau_2^2 \Gamma_2^{\dagger b} \Gamma_2^f e^{-i \chi} + \tau_1^2
\Gamma_1^{\dagger b} \Gamma_1^f e^{-i \chi}+\mathcal{O}(\tau^3), \nonumber \\
T'&=& i \tau_1 \left\lbrace \Gamma_1^{b} e^{i \chi/2} -
\Gamma_1^{f} e^{-i \chi/2} \right\rbrace \\
& & +\, i \tau_2 e^{i \Phi} \nonumber
\left\lbrace \Gamma_2^{b} e^{i \chi/2} - \Gamma_2^{f} e^{-i \chi/2}
\right\rbrace+\mathcal{O}(\tau^2), \\
T&=&i\tau_1 \left\lbrace \Gamma_1^{\dagger b} e^{-i \chi/2} -
\Gamma_1^{\dagger f} e^{i \chi/2} \right\rbrace \\
  & & +\, i\tau_2 e^{-i \Phi} \left\lbrace \Gamma_2^{\dagger b} e^{-i \chi/2} -
\Gamma_2^{\dagger f} e^{i \chi/2} \right\rbrace +\mathcal{O}(\tau^2), \nonumber \\
 R'&=&1-\tau_1^2-\tau_2^2 -\tau_1 \tau_2 \left\lbrace
\Gamma_1^{b} \Gamma_2^{\dagger b} e^{- i \Phi} + \Gamma_2^{f} \Gamma_1^{\dagger
f} e^{i \Phi}\right\rbrace \nonumber\\
    & & +\, \tau_2^2 \Gamma_2^{b} \Gamma_2^{\dagger f} e^{i \chi} + \tau_1^2
\Gamma_1^{b} \Gamma_1^{\dagger f} e^{i \chi} \\
    & & +\,\tau_1 \tau_2 \left\lbrace \Gamma_2^{b} \Gamma_1^{\dagger f} e^{i \Phi}
+ \Gamma_1^{b} \Gamma_2^{\dagger f} e^{-i \Phi}\right\rbrace  e^{i
\chi}+\mathcal{O}(\tau^3).\nonumber 
\end{eqnarray}
They can be easily deduced using the set of rules formulated in section II.C.

\section{Tunneling action}

In this Appendix we present technical details of our calculations leading
to the tunneling action~(\ref{eq:AES_action}). The first step is
to evaluate 
the trace in the expansion~(\ref{eq:Trexpand}) of the functional determinant 
in the channel and time domain using the
explicit form of the $Q$-matrix.  A representation of the matrix $Q$ 
in terms of the ``hopping'' operators $\Gamma_i$  
defined by Eq.~(\ref{eq:Gamma}) is given in Appendix B. After straightforward 
(albeit lengthy) calculations, we get
\begin{eqnarray}
\label{eq:ActionT}
 i\mathcal{A}_t & = &\sum_{ij}\tau_i\tau_j \int dt_{1,2}\sum_{aa'}
 P_{aa'} 
\\ &\times& 
e^{-i\theta^{a}_i(t_1)}\bar{\Pi}_{ij}^{aa'}(t_1,t_2)e^{
i{\theta}^{a'}_j(t_2)}. \nonumber
\end{eqnarray}
Here $P_{aa'}=1$ if $a=a'$ and $P_{aa'}=-1$ otherwise ($a$ is the
Keldysh index). Further,
\begin{equation}
{\theta}^{a}_i(t)={\theta}^{+,a}(x_i^+,t)-{\theta}^{-,a}(x_i^-,t)
\end{equation}
is the relative phase expressed in terms of 
the phases ${\theta}^{\pm,a}$ accumulated by electron along the paths 
from scattering point $x^\pm_i$ to the drain at upper and lower arms, 
\begin{equation}
{\theta}^{\pm,a}(x^{\pm},t)=-\frac{1}{v}\int_{x^{\pm}}^{L^{\pm}}
dx' \varphi^{\pm}_a(x',t+(x'-x)/v).
\label{eq:Kin}
\end{equation}
The polarization operator $\bar{\Pi}_{ij}^{aa'}$ has the same
structure as the operator  
${\Pi}_{ij}^{aa'}$, defined by the relations~(\ref{eq:Pol1}),
(\ref{eq:Pol2}) and (\ref{eq:Pol3}). 
The difference is that $\bar{\Pi}_{ij}^{aa'}$ is built up from the
gauge transformed distribution 
function $\bar f(t,t')$, given by Eq.~(\ref{eq:f-gauge-transformed}). 
We also note
that the phases ${\theta}^{\pm,a}$ are 
expressed via $\varphi^{\pm,a}$ by the kinematic relation (\ref{eq:Kin}).

The next step is to show that the action~(\ref{eq:ActionT}), in fact,
coincides with its final  
form~(\ref{eq:AES_action}). For that we use the relation
\begin{equation}
 D_0^{R/A}(x,t)=\pm v^{-1} \theta(\pm x)\delta(t-x/v),
 \label{eq:D0_RA}
\end{equation}
which follows directly from the definitions~(\ref{eq:D0}) and the
standard identity  of the Keldysh theory
\begin{equation}
D_0^{R/A}(x,t)= \theta(\pm t) \left(D_0^{>}(x,t) -D_0^{<}(x,t) \right).  
\end{equation}
The kinematic phase ${\theta}^{\pm,a}$ can be now represented in the
symbolic form as 
\begin{equation}
{\theta}^{\pm,a}(x,t) = \left( D_0^A \varphi^{\pm}_a\right)(x,t).
\label{eq:theta_small}
\end{equation}
To proceed further we express the matrix elements of the polarization
operator  $\bar{\Pi}_{ij}^{aa'}$ 
in terms of the original distribution functions
$f_\alpha^{\pm}(t,t')$, using Eq.~(\ref{eq:lambda}). 
This transformation reduces the action~(\ref{eq:ActionT}) to the
action (\ref{eq:AES_action}), with
the phases $\Theta^{\pm,a}$  equal to 
\begin{equation}
 \Theta^{\pm,a}(x,t) = \lambda^\pm(t-x/v)-\theta^{\pm,a}(x,t).
 \label{eq:Theta_large}
\end{equation}
Let us now show that the relation (\ref{eq:Theta_large}) is in fact 
equivalent to Eq.~(\ref{eq:Theta}).

First, using the explicit form~(\ref{eq:D0_RA}) of the retarded and
advanced particle-hole propagator, one can express  
the phases $\psi^\pm_{f/b}$, defined by Eq.~(\ref{eq:psi}),  as
\begin{equation}
\psi^{\pm}_{b/f} = \left( (D_0^A-D_0^R)\, \varphi^{\pm}_{b/f}\right)(x,t).
\end{equation}
We also use the fact that at zero temperature the
Fermi and the Bose distribution functions are closely related to each
other, namely $f_F(t-t')=-n_B(t-t')$. 
This enables us to write the gauge phase~(\ref{eq:lambda}) as
\begin{eqnarray}
\lambda^\pm(x-v/t)&=&
\left((n_B+1)(D_0^A-D_0^R)\,\varphi^{\pm}_f\right)(x,t) \nonumber\\ 
&-& \Bigl( n_B(D_0^A-D_0^R)\,\varphi^{\pm}_b\Bigr)(x,t).
\end{eqnarray}
From this relation and Eqs.~(\ref{eq:theta_small}) and
(\ref{eq:Theta_large}) one can finally see that 
\begin{eqnarray}
\Theta^{\pm,f}=- D_0^T \varphi_f^\pm + D_0^{<} \varphi_b^\pm, \\
\Theta^{\pm,b}=- D_0^> \varphi_f^\pm + D_0^{\tilde{T}} \varphi_b^\pm,
\end{eqnarray}
which agrees with the desired relation~(\ref{eq:Theta}). To obtain
these identities, we have 
used the standard properties of the Keldysh propagators, 
\begin{eqnarray}
D_0^{<} &=& n_B ( D_0^R - D_0^A), \quad D_0^{>} = (n_B +1 )( D_0^R -
D_0^A), \nonumber \\ 
D_0^T &=& D_0^R + D_0^<, \quad D_0^{\tilde T} = D_0^< - D_0^A.
\end{eqnarray}
This completes our proof of equivalence between two forms of the 
action, Eqs.~(\ref{eq:ActionT}) and  
(\ref{eq:AES_action}).

\section{Stationary phase method}

To describe the oscillations of the plasmon correlation function $J^>_P(x,t)$,
we analyze the integral~(\ref{eq:PlasmonIntegral}) in the short time
limit $t\ll x/v$. 
Then the optimal momentum is small, $q_* \ll b^{-1}$, which enables us to write
\begin{equation}
\phi(q)=\phi_* g(q/q_*), 
\end{equation}
where $|\phi_*|\gg 1$ is the stationary phase and 
\begin{equation}
g(\lambda)=\lambda (1-\ln\lambda).
\end{equation}
Let us consider 
\begin{equation}
  \partial_{\phi_*} J^>_{P}=-i\int_0^{\infty} d\lambda(1-\ln \lambda)
e^{i\phi_* g(\lambda)} e^{-a q_* \lambda}.
\end{equation}
The main contributions to this oscillatory integral come from the stationary
points and from the end points. Here we concentrate on the stationary phase
contribution, which is responsible for the oscillations in the correlation
function. Thus we get
\begin{equation}
 \label{eq:Jphi}
 \partial_{\phi_*} J^{>}_{P}=-i\sqrt{\frac{2\pi i}{-\phi_*}} e^{i\phi_*}.
\end{equation}
Integrating back this equation and taking into account the boundary condition,
$\mathrm{Im} J^>_{P}\rightarrow \pi/2$ for $t\rightarrow \infty$
($\phi_*\rightarrow-\infty$), one arrives at
\begin{equation}
\label{approximate_Im_J}
 \mathrm{Im} J^{>}_{P}=\frac{\pi}{2}-\sqrt{2} \pi {\rm Re}\,
\mathrm{Erfc} (\sqrt{-i\phi_*}),
\end{equation}
where ${\rm Erfc}(z)$ is the complementary error function.
This formula does not apply at $t\rightarrow 0$ ($\phi_*\rightarrow
0$), where the stationary phase is small and the associated
approximation breaks down. As is seen in Fig.~\ref{fig:Jp},
at $\phi_*\sim 1$ the oscillatory behavior crosses over into a smooth
one, and with further decrease of $t$ the function $\mathrm{Im}
J^{>}_{P}$ saturates at $-\pi/2$.   
We can approximate this crossover at $\phi_*\sim 1$ and saturation
at $\phi_*\sim 0$  by
replacing $\phi_*$  in Eq.~(\ref{approximate_Im_J})
by $\phi_*-\phi_0$, where the phase $\phi_0\approx0.12$ 
satisfies the equation 
\begin{equation}
\mathrm{Re}\,
\mathrm{Erfc}(\sqrt{i \phi_0})=1/\sqrt{2}.
\end{equation}
To improve further the accuracy of the approximation, one can also replace the
limiting expression for  
the stationary phase~(\ref{eq:phi_star}) (which was found using the
logarithmic approximation 
to the plasmon dispersion relation) by its exact value found from the
numerical solution of Eq.~(\ref{eq:st_phase}). This yields the final approximation, 
which is shown in Fig.\ref{fig:Jp} by the dotted line.

\end{document}